\documentclass[apjl]{myemulateapj}

\usepackage{comment}
\usepackage{ifthen}

\newboolean{emulateapj}
\setboolean{emulateapj}{true}

\newcommand{\ctbd}[1]{}

\newcommand{\lc}{light curve}
\newcommand{\lcs}{light curves}
\newcommand{\Lc}{Light curve}

\newcommand{\kms}{\ensuremath{\rm km\,s^{-1}}}
\newcommand{\ms}{\ensuremath{\rm m\,s^{-1}}}

\newcommand{\gcmc}{\ensuremath{\rm g\,cm^{-3}}}

\newcommand{\rsun}{\ensuremath{R_\sun}}
\newcommand{\msun}{\ensuremath{M_\sun}}
\newcommand{\lsun}{\ensuremath{L_\sun}}

\newcommand{\rstar}{\ensuremath{R_\star}}
\newcommand{\mstar}{\ensuremath{M_\star}}

\newcommand{\teffstar}{\ensuremath{T_{\rm eff}}}

\newcommand{\rjup}{\ensuremath{R_{\rm J}}}
\newcommand{\mjup}{\ensuremath{M_{\rm J}}}

\newcommand{\flwof}{\mbox{FLWO 1.2 m}}

\newcommand{\band}[1]{\ensuremath{#1}-band}

\newcommand{\hatcur}{HAT-P-8}
\newcommand{\hatcurb}{HAT-P-8b}

\newcommand{\hatcurCCra}{\ensuremath{22^{\mathrm{h}}52^{\mathrm{m}}09^{\mathrm{s}}.85}}	%
\newcommand{\hatcurCCdec}{\ensuremath{+35^{\circ}26'49''.5}}		%
\newcommand{\hatcurCCmag}{\ensuremath{9.68}}		
\newcommand{\hatcurCCtwomass}{2MASS~22520985+3526495}	
\newcommand{\hatcurCCgsc}{GSC~02757-01152}		
\newcommand{\hatcurCCtassmv}{10.30}			
\newcommand{\hatcurCCtassvi}{\ensuremath{0.62\pm0.09}}	

\newcommand{\hatcurLCdip}{\ensuremath{7.0}}				
\newcommand{\hatcurLCrprstar}{\ensuremath{0.0953\pm0.0009}}		
\newcommand{\hatcurLCimp}{\ensuremath{0.32^{+0.09}_{-0.19}}}		
\newcommand{\hatcurLCdur}{\ensuremath{0.1587\pm0.0011}}			
\newcommand{\hatcurLCingdur}{\ensuremath{0.0144\pm 0.0010}}		
\newcommand{\hatcurLCP}{\ensuremath{3.076320\pm0.000004}}		
\newcommand{\hatcurLCPprec}{\ensuremath{3.076320}}			
\newcommand{\hatcurLCPshort}{3.0763}					
\newcommand{\hatcurLCMT}{\ensuremath{54,437.67582\pm0.00034}}		

\newcommand{\hatcurSMEteff}{\ensuremath{6200\pm80}}			%
\newcommand{\hatcurSMEzfeh}{\ensuremath{+0.01\pm0.08}}			%
\newcommand{\hatcurSMEvsin}{\ensuremath{11.5\pm0.5}}			%

\newcommand{\hatcurYYm}{\ensuremath{1.28\pm0.04}}			%
\newcommand{\hatcurYYr}{\ensuremath{1.58^{+0.08}_{-0.06}}}		%
\newcommand{\hatcurYYlogg}{\ensuremath{4.15\pm0.03}}			%
\newcommand{\hatcurYYlum}{\ensuremath{3.3^{+0.4}_{-0.3}}}		%
\newcommand{\hatcurYYmv}{\ensuremath{3.48\pm0.12}}			%
\newcommand{\hatcurYYvi}{\ensuremath{0.58\pm0.02}}			%
\newcommand{\hatcurYYage}{\ensuremath{3.4\pm1.0}}			%

\newcommand{\hatcurRVK}{\ensuremath{153.1\pm3.9}}			%

\newcommand{\hatcurPPi}{\ensuremath{87\fdg5^{+1.9}_{-0.9}}}		%
\newcommand{\hatcurPPlogg}{\ensuremath{3.23\pm0.03}}			%
\newcommand{\hatcurPPar}{\ensuremath{6.35^{+0.34}_{-0.17}}}		%
\newcommand{\hatcurPParel}{\ensuremath{0.0487\pm0.0026}}		%
\newcommand{\hatcurPPrho}{\ensuremath{0.568\pm0.048}}			%

\newcommand{\hatcurPPm}{\ensuremath{1.52^{+0.18}_{-0.16}}}		%
\newcommand{\hatcurPPmshort}{\ensuremath{1.52}}				%
\newcommand{\hatcurPPr}{\ensuremath{1.50^{+0.08}_{-0.06}}}		%
\newcommand{\hatcurPPmrcorr}{\ensuremath{0.77}}				%
\newcommand{\hatcurPPteff}{\ensuremath{1700\pm35}}			%
\newcommand{\hatcurPPtheta}{\ensuremath{0.061\pm0.003}}

\newcommand{\hatcurXdist}{\ensuremath{230\pm15}}			%

\newcommand{\df}{\dotfill}
\newcommand{\p}{$\pm$}
\newcommand{\nd}{\nodata}

\ifthenelse{\boolean{emulateapj}}{
	\newcommand{\titledag}{$\dagger$}
}{
	\newcommand{\titledag}{\dagger}
}

\shortauthors{Latham et al.}
\shorttitle{\hatcur{}}

\begin{document}


\title{HATNet Field G205: Follow-Up Observations of 28
Transiting-Planet candidates and Confirmation of the Planet 
\hatcurb\altaffilmark{\titledag}}

\altaffiltext{$\dagger$}
{Based in part on observations obtained at the W.~M.~Keck
Observatory, which is operated by the University of California and
the California Institute of Technology. Keck time has been
granted by NOAO (A285Hr).}


\author{
	David~W.~Latham\altaffilmark{1},
	G\'asp\'ar~\'A.~Bakos\altaffilmark{1,2},
	Guillermo~Torres\altaffilmark{1},
	Robert~P.~Stefanik\altaffilmark{1},
	Robert~W.~Noyes\altaffilmark{1},
	G\'eza~Kov\'acs\altaffilmark{3},
	Andr\'as~P\'al\altaffilmark{1,4},
	Geoffrey~W.~Marcy\altaffilmark{5},
	Debra~A.~Fischer\altaffilmark{6},
	R.~Paul~Butler\altaffilmark{7},
	Brigitta~Sip\H{o}cz\altaffilmark{4,1},
	Dimitar~D.~Sasselov\altaffilmark{1},
	Gilbert~A.~Esquerdo\altaffilmark{1},
	Steven~S.~Vogt\altaffilmark{8},
	Joel~D.~Hartman\altaffilmark{1},
	G\'abor~Kov\'acs\altaffilmark{1},
	J\'ozsef~L\'az\'ar\altaffilmark{9},
	Istv\'an~Papp\altaffilmark{9},
	P\'al~S\'ari\altaffilmark{9}
}

\altaffiltext{1}{Harvard-Smithsonian Center for Astrophysics,
	60 Garden Street, Cambridge, MA 02138}

\altaffiltext{2}{NSF Fellow}

\altaffiltext{3}{Konkoly Observatory, Budapest, Hungary}

\altaffiltext{4}{Department of Astronomy,
	E\"otv\"os Lor\'and University, Budapest, Hungary.}

\altaffiltext{5}{Department of Astronomy, University of California,
	Berkeley, CA 94720}

\altaffiltext{6}{Department of Physics and Astronomy, San Francisco
	State University, San Francisco, CA 94132}

\altaffiltext{7}{Department of Terrestrial Magnetism, Carnegie
        Institution of Washington, 5241 Broad Branch Road NW, Washington,
        D.C. 20015-1305} 

\altaffiltext{8}{University of California Observatories/Lick Observatory, University of California at 
        Santa Cruz, Santa Cruz, CA 95064}

\altaffiltext{9}{Hungarian Astronomical Association, Budapest, 
	Hungary}


\begin{abstract} 
We report the identification of 32 transiting-planet candidates in
HATNet field G205.  We describe the procedures that we have used to
follow up these candidates with spectroscopic and photometric
observations, and we present a status report on our interpretation of
the 28 candidates for which we have follow-up observations.
Eight are eclipsing binaries with orbital solutions whose
periods are consistent with their photometric ephemerides; two of
these spectroscopic orbits are singled-lined and six are double-lined.
For one of the candidates, a nearby but fainter eclipsing binary
proved to be the source for the HATNet light curve, due to blending in
the HATNet images.  Four of the candidates were found to be rotating more
rapidly than $v \sin i = 50$ \kms\ and were not pursued further. Thirteen
of the candidates showed no significant velocity variation at the level of
0.5 to 1.0 \kms.  Seven of these were eventually withdrawn as photometric
false alarms based on an independent reanalysis using more sophisticated
tools.  Of the remaining six, one was put aside because a close visual
companion proved to be a spectroscopic binary, and two were not followed
up because the host stars were judged to be too large.  Two of the remaining
candidates are members of a visual binary, one of which was previously
confirmed as the first HATNet transiting planet, HAT-P-1b.  In this paper
we confirm that the last of this set of candidates is also a a transiting
planet, which we designate \hatcur{b}, with mass $M_{\rm p} = \hatcurPPm\,\mjup$,
radius $R_{\rm p} = \hatcurPPr\,\rjup$, and photometric period $P = \hatcurLCP$
days. \hatcur{b} has an inflated radius for its mass, and a large mass
for its period.  The host star is a solar-metallicity F dwarf, with
mass $\mstar = \hatcurYYm\,\msun$ and $\rstar = \hatcurYYr
\rsun$.

\end{abstract}






\section{INTRODUCTION}
\label{sec:introduction}

Photometric surveys are now finding many hundreds of stars that
exhibit periodic dimmings that look like they might be due to
transiting planets, but so far the vast majority of these candidates
have turned out to be stellar systems involving eclipsing binaries and
not planets.  To make a convincing case that the light curve is due to
a transiting planet it is necessary, but not sufficient, to show that
the detailed shape of the dimming, including depth and duration, can
be accurately modeled as a planet.  The proof of the pudding is then
an orbit for the host star that matches the photometric ephemeris and
implies a planetary mass for the companion.  For a spectroscopic orbit
it is also necessary to demonstrate that the measured velocity variations
are not due to astrophysical effects other than orbital motion.

As of 1 July 2008, spectroscopic orbits had been properly published
for 30 stars hosting transiting planets.  In five cases (HD 209458, HD
149026, HD 189733, Gliese 436, and HD 17156) the orbit came first from
Doppler surveys, and transits were subsequently identified.  These
five have been the source of the most interesting exoplanet
astrophysics that has emerged so far, primarily because they are the
nearest and brightest.  In contrast, the initial excitement of
confirming a few planets from the OGLE photometric surveys has now
cooled because of the difficulty of making follow-up observations for
such faint systems.  Wide-angle ground-based photometric surveys
utilizing small telescopes are now the main source of new transiting
planets, with 19 confirmed and published planets from TrES, HAT, XO,
and WASP, and many more on the way.  Although generally fainter than
the five from Doppler surveys, some of these are bright enough for
interesting follow-up work, and the growing numbers are fleshing out
the extraordinary population of close-in giant planets.

At the Harvard-Smithsonian Center for Astrophysics (CfA), some of us
have been following up transiting planet candidates for almost ten
years, starting with targets supplied by the Vulcan team in 1999.  We
soon learned that the sample was dominated by systems involving
eclipsing binaries \citep{Latham:03}.  Indeed, we have not yet been
able to confirm that any of the 66 Vulcan candidates that we observed
are actually planets.  More recently, two other wide-angle surveys,
TrES and HATNet, have been the source of a large number of
transiting-planet candidates that have been followed up initially at
CfA.  As of 1 July 2008 a total of 811 candidates from these two
surveys had been observed or scheduled for observations, but as of 1
July 2008 only 11 of these had led to proper publication as confirmed
planets.

Over the years our procedures for reducing and analyzing HATNet
photometry and for identifying good transiting-planet candidates have
evolved considerably, and we have encountered a variety of
stellar systems that can mimic transiting planets.  Thus one of the
goals of this paper is to report our present procedures and some of
the lessons learned.  A second, and perhaps more important goal is to
document the false positives that we have identified in the process of
our work to follow up candidates identified in HATNet field G205.  As
more of the sky gets covered by wide-angle surveys, it is inevitable
that some of the same candidates will be identified by more than one
survey.  Ways must be found to document the follow-up work that has
already been done, to avoid wasteful duplication of effort.  Last, but
not least, we confirm and characterize \hatcur{b}, a second planet to
join with HAT-P-1b in field G205.


\section{PHOTOMETRIC DETECTION}
\label{sec:detection}

The HATNet telescopes \mbox{HAT-5}, \mbox{HAT-6}, \mbox{HAT-8} and
\mbox{HAT-9} \citep[HATNet;][]{Bakos:02, Bakos:04} observed HATNet
field G205, centered at $\alpha = 22^{\rm h} 56^{\rm m}$, $\delta =
+37\arcdeg 30\arcmin$, on a nightly basis between 2003 September 29
and 2004 February 1, and between 2006 July 3 and 2006 July
24. Exposures of 5 minutes were obtained at a 5.5-minute cadence
whenever conditions permitted; all in all 4460 exposures were secured,
each yielding photometric measurements for approximately $45,000$
stars in the field down to $I\sim14.0$ mag. The field was observed in
network mode, exploiting the longitude separation between
\mbox{HAT-5/6}, stationed at the Smithsonian Astrophysical
Observatory's (SAO) Fred Lawrence Whipple Observatory (FLWO) on Mount
Hopkins in Arizona ($\lambda=111\arcdeg$W), and \mbox{HAT-8/9},
installed on the rooftop of SAO's Submillimeter Array building
atop Mauna Kea, Hawaii ($\lambda=155\arcdeg$W).

At the time of the early reduction of HATNet frames and the initial
search for planet candidates, starting in the Fall of 2003, our tools
and strategies were much less developed than now.  Although the Box
Least Squares \citep[BLS;][]{Kovacs:02} algorithm was already in
routine use to search for periodic transit-like dips, the light curves
themselves were still plagued by a variety of subtle trends. Some of
these trends were not understood, and our intense effort to suppress
them led to the development of the Trend Filtering Algorithm
\citep[TFA;][]{Kovacs:05}.  Later we gained a better understanding of
the trends, and some were associated with ``external parameters'',
such as sub-pixel position, airmass, stellar profile parameters, and
the topology of other objects around the selected star. This led to
the development of the External Parameter Decorrelation technique
\citep[EPD, described briefly in][]{Bakos:07b}, whereby we first
correct for trends that have known underlying parameters. The EPD
corrected \lcs\ are then processed with TFA.

In the early reductions, these tools were simply not available. Thus, a
variety of effects such as subtle out-of-transit variations or
color-dependent trends with atmospheric extinction went unrecognized or
uncorrected, leading to i) undetected shallow transits, and ii) false
alarms due to remaining trends.  Furthermore, we were anxious to
confirm our first HATNet transiting planet, and there was a strong
temptation to retain as many candidates as possible for follow-up
observations, including some where the detection of transit events was
marginal. Field ``G205'' reductions witnessed all these changes, and
the resulting candidate list is thus somewhat inhomogeneous.

The current pipeline, using multi-aperture photometry, EPD, TFA, and
careful multi-step selection procedures, allows us to detect transit
events for fainter stars.  In addition, our rate of photometric false
alarms has declined, thanks both to the improved photometry and several
years of additional experience in how to evaluate the reliability of a
possible detection.

Periodic transit-like dips in brightness were identified
initially for 31 HATNet targets in field G205.  These candidates are
listed in Table~\ref{tab:candidates}, where we provide the internal
HATNet candidate identification (HTR stands for HATNet TRansit
candidate), position, estimated $V$ magnitude, proper motion in mas yr$^{-1}$,
and $J-K_{\rm s}$ color from the Two Micron
All-Sky Survey \citep[2MASS;][]{Skrutskie:06}.
The periods that are given to two decimal places are from the original
analysis of the light curves for periodic transits, while the periods
that are accompanied by errors and epochs are from a reanalysis using
the most recent tools.  In the original photometric reduction of the
HATNet G205 field, HTR205-001 yielded a light curve and was also
identified as a planet candidate.  The standard photometric reduction
now relies on astrometry from 2MASS.  HTR205-001 is flagged as having
poor quality in 2MASS, and therefore it was not included in the most
recent photometric reduction.  For ten of the original candidates
(HTR205-002, HTR205-005, HTR205-006, HTR205-008, HTR205-016, HTR205-019, HTR205-025,
HTR205-026, HTR205-028, and HTR205-030), the latest reductions and analysis no longer
yield reliable periods.  They are now considered photometric
false alarms and are also listed with their original periods to two
decimal places.  For six of the candidates (HTR205-004, HTR205-010,
HTR205-011, HTR205-012, HTR205-017, and HTR205-020) the periods listed
in Table~\ref{tab:candidates} have been doubled from the values
yielded by the most recent analysis of the light curves in order to
closely match the periods yielded by the spectroscopic orbits reported in this
paper.  The ephemerides listed for HTR205-023 and HTR205-024 are taken
from \citet{Winn:07} and from the solution reported later in this paper,
respectively.

One of the initial candidates, HTR205-007, turned out to be a close
pair of stars separated by only $3\farcs2$, so it is listed as two
candidates in Table~\ref{tab:candidates}, HTR205-007E and HTR205-007W.
The photometry of another one of the initial candidates, HTR205-018,
turned out to be contaminated by the light of a nearby eclipsing
system.  This problem was revealed by an additional step in the
photometric analysis, which is now applied routinely to all
candidates.  To explore the possibility that a transiting-planet
candidate is a blend with a nearby variable, we now inspect the light
curves of all the stars within about $1 \arcmin$ of the candidate and
brighter than $I = 14$ mag.  In the case of HTR205-018, a star
separated by $22\arcsec$ from HTR205-018 (at $\alpha =
22^{\mathrm{h}} 48^{\mathrm{m}} 23^{\mathrm{s}}.4, ~\delta =
+35\arcdeg 55\arcmin 5\arcsec$) and approximately 1.2 mag fainter in
the HATNet band, showed events with the same ephemeris, but with a
depth of 0.0199 mag compared to 0.007 mag for HTR205-018.  Correcting
for the dilution by the contaminating light from HTR205-018, the
actual depth works out to about 0.05 mag.  This is too deep to be a
transiting planet around HTR205-018 if it is a Sun-like star, as implied by
its $J-K = 0.35$ color index, so both the companion and HTR205-018 were 
withdrawn as viable candidates.

In a previous publication \citep{Bakos:07a}, HTR205-023 was announced
as HAT-P-1b, the first confirmed transiting planet from HATNet.  In
this paper we confirm that a second candidate in field G205,
HTR205-024, is also a transiting planet, which we designate
\hatcur{b}.

\ifthenelse{\boolean{emulateapj}}{
	\begin{deluxetable*}{lcrccll}
}{
	\begin{deluxetable}{lcrccll}
}
\tabletypesize{\scriptsize}
\tablewidth{0pc}
\tablenum{1}
\tablecaption{Candidates from HATNet Field G205}
\label{tab:candidates}
\tablehead{
\colhead{~~~~~Star~~~~~~~~~~} &
\colhead{RA (2000) Dec} &
\colhead{$V$} &
\colhead{PM} &
\colhead{$J-K_{\rm s}$}&
\colhead{$P$}&
\colhead{Epoch (HJD)}}
\startdata
HTR205-001  \df &  22:35:31.8  +39:53:59 & 11.03 & 12.1 & 0.08 & 1.25                 & \nd           \\
HTR205-002  \df &  22:37:12.3  +38:13:43 & 10.06 & 8.52 & 0.63 & 2.44                 & \nd           \\
HTR205-003  \df &  22:37:36.4  +34:36:24 & 12.48 & 24.7 & 0.33 & 2.17925  \p 0.00025  & 24452950.3206 \\
HTR205-004  \df &  22:37:39.5  +38:07:43 & 12.22 & 27.6 & 0.42 & 5.45498  \p 0.00045  & 24452950.0516 \\
HTR205-005  \df &  22:40:06.2  +37:31:39 & 11.88 & 5.81 & 0.39 & 4.35                 & \nd           \\
HTR205-006  \df &  22:40:59.5  +35:25:22 & 11.82 & 1.07 & 0.76 & 3.61                 & \nd           \\
HTR205-007W \df &  22:44:02.4  +40:05:11 & 12.08 & 33.0 & 0.35 & 1.88249  \p 0.00032  & 24452911.5222 \\
HTR205-007E \df &  22:44:02.6  +40:05:11 & 11.50 & 33.0 & 0.35 & 1.88249  \p 0.00032  & 24452911.5222 \\
HTR205-008  \df &  22:44:49.4  +39:28:50 & 10.61 & 10.6 & 0.78 & 1.63                 & \nd           \\
HTR205-009  \df &  22:54:56.5  +34:31:42 & 10.93 & 6.04 & 0.27 & 1.55845  \p 0.00018  & 24452911.4381 \\
HTR205-010  \df &  22:56:01.7  +37:51:06 & 11.33 & 7.02 & 0.31 & 2.51441  \p 0.00041  & 24452912.8156 \\
HTR205-011  \df &  23:01:47.4  +38:06:22 & 11.19 & 13.1 & 0.38 & 3.13933  \p 0.00029  & 24452914.2476 \\
HTR205-012  \df &  23:06:23.6  +39:07:19 & 12.28 & 18.4 & 0.35 & 3.15930  \p 0.00060  & 24452910.9429 \\
HTR205-013  \df &  23:08:08.3  +33:38:03 & 10.72 & 24.7 & 0.28 & 2.23074  \p 0.00026  & 24452912.7566 \\
HTR205-014  \df &  23:09:09.5  +36:40:38 & 10.28 & 10.7 & 0.13 & 7.20626  \p 0.00071  & 24452916.2177 \\
HTR205-015  \df &  23:12:13.3  +37:44:09 & 11.11 & 11.1 & 0.64 & 1.192870 \p 0.000090 & 24452910.9437 \\
HTR205-016  \df &  22:53:11.2  +34:44:41 & 11.21 & 10.1 & 0.81 & 1.56                 & \nd           \\
HTR205-017  \df &  22:41:32.2  +36:36:22 & 11.94 & 2.76 & 0.32 & 4.79886  \p 0.00055  & 24452914.8199 \\
HTR205-018  \df &  22:48:21.8  +35:54:53 & 11.71 & 26.9 & 0.35 & 0.55315  \p 0.00003  & 24452910.8027 \\
HTR205-019  \df &  22:57:54.9  +33:39:00 & 11.47 & 9.79 & 0.61 & 5.14                 & \nd           \\
HTR205-020  \df &  22:41:46.7  +36:49:28 & 10.57 & 7.43 & 0.09 & 6.52295  \p 0.00082  & 24452913.7491 \\
HTR205-021  \df &  22:36:08.1  +37:50:26 & 10.27 & 86.1 & 0.40 & 0.789765 \p 0.000039 & 24452950.0783 \\
HTR205-022  \df &  22:57:45.9  +38:40:26 & 10.00 & 49.5 & 0.27 & 4.46541  \p 0.00046  & 24452912.7253 \\
HTR205-023  \df &  22:57:46.8  +38:40:29 & 10.40 & \nd  & 0.30 & 4.46543  \p 0.00049  & 24453997.7926 \\
HTR205-024  \df &  22:52:09.8  +35:26:49 & 10.17 & 76.0 & 0.26 & 3.07632  \p 0.00045  & 24454437.6758 \\
HTR205-025  \df &  22:36:44.5  +40:03:13 & 10.93 & 2.2  & 0.78 & 6.39                 & \nd           \\
HTR205-026  \df &  22:36:45.1  +35:08:06 & 12.09 & 20.5 & 0.36 & 3.73                 & \nd           \\
HTR205-027  \df &  22:42:07.5  +39:02:43 & 13.63 & 58   & 0.89 & 0.555000 \p 0.000012 & 24452950.0867 \\
HTR205-028  \df &  22:45:51.6  +40:28:37 & 13.92 & 12.4 & 0.63 & 1.56                 & \nd           \\
HTR205-029  \df &  23:10:07.1  +41:05:11 & 13.51 & \nd  & 0.39 & 3.72630  \p 0.00039  & 24453598.0809 \\
HTR205-030  \df &  22:41:46.2  +37:13:43 & 12.39 & 33.7 & 0.39 & 1.60                 & \nd           \\
HTR205-031  \df &  23:06:52.7  +34:21:24 & 13.02 &  1.2 & 0.29 & 0.750731 \p 0.000051 & 24452910.9452
\enddata
\tablecomments{
column (1): HATNet transiting-planet candidate designation, \\
column (2): Right Ascension and Declination from HATNet, \\
column (3): Estimated $V$ magnitude, \\
column (4): Proper Motion, mas yr$^{-1}$, \\
column (5): $J-K_{\rm s}$ color from 2MASS, \\
column (6): Photometric period, days, \\
column (7): Epoch of mid transit, Heliocentric Julian Day.}
\ifthenelse{\boolean{emulateapj}}{
	\end{deluxetable*}
}{
	\end{deluxetable}
}


\section{FOLLOW-UP OBSERVATIONS}
\label{sec:followup}

Our strategy for following up transiting-planet candidates identified
by wide-angle ground-based photometric surveys such as HATNet involves
both spectroscopy and photometry.  The discovery light curves from
small cameras such as those used by HATNet rarely provide the
photometric precision and time resolution needed for accurate
determinations of planetary radii.  On the other hand, large-format CCD
cameras on wide-field meter-class telescopes are well suited for
obtaining high-quality light curves for candidates identified by
wide-angle surveys such as HATNet.  The image field needs to be wide
enough to provide a good selection of comparison stars for the
differential photometry, while the telescope aperture needs to be
large enough to receive enough flux to reach optimal signal-to-noise
and at the same time keep the exposure times short enough for good time
resolution at the faint end, typically 13th magnitude.  At the bright
end, if the telescope aperture is too large the exposure times end up
being too short because of detector saturation.  This leads to a poor
duty cycle because of the fixed detector overhead time, and
scintillation noise can dominate the photometric performance.  To some
extent the saturation problem is alleviated if we use the less
sensitive SDSS $z$ band in the near infrared; however, the primary reason
for using the $z$ band is that stellar limb darkening is less
important at longer wavelengths, and the effects of atmospheric
extinction and scintillation also get smaller.  If the star is
nevertheless so bright that saturation is a potential problem, we
defocus the telescope unless the field near the target is very
crowded.  For the photometric follow-up of many of the most
interesting candidates identified by HATNet we have used KeplerCam
on the 1.2-m reflector at FLWO. KeplerCam uses a 4Kx4K Fairchild 486
CCD with $0\farcs33$ pixels, $23 \arcmin$ field of view, and camera
overhead of 12 seconds for 2x2
 binning.

However, photometric observations with KeplerCam are not usually the
first step in the follow-up of a HATNet candidate, for the simple
reason that observations during transit occupy only a few percent of
the time and are difficult to schedule.  Instead our usual strategy
has been to start with a spectroscopic reconnaissance, to look for
orbital motion due to a stellar companion and to better characterize
the target.  These observations are much easier to schedule, because
the radial velocity varies continuously throughout an orbit.

For the reconnaissance spectroscopy we have been using the CfA Digital
Speedometers \citep{Latham:92}, mostly on the 1.5-m Tillinghast
Reflector at FLWO, but also on the 1.5-m Wyeth Reflector at the Oak
Ridge Observatory in the Town of Harvard, Massachusetts, up until 2005
when that facility was terminated.  These venerable instruments use
intensified photon-Counting Reticon detectors on identical echelle
spectrographs to record 45\AA\ of spectrum centered at 5187\AA, thus
including the gravity-sensitive Mg b features.  Our usual strategy for
follow-up of a new candidate has been to obtain an initial spectrum.
If it shows decent lines we then get a second spectrum, the next night
if possible, so that we can check to see if the velocity varies.  For
slowly-rotating solar-type stars the typical velocity precision is 0.5
\kms, which is adequate for the detection of a companion down to a
limit of about 10 \mjup\ for periods of a few days.  Stellar
companions induce much larger orbital amplitudes and are easy to
identify if there are only two objects in the system.

The same spectra that are used to determine velocities are also used
to characterize the effective temperature, surface gravity, and
projected rotational velocity of the host star, using an extensive
library of synthetic spectra to find the template spectrum that gives
the best match to the observed spectra, based on grids of
one-dimensional correlations.  Because of the narrow wavelength range
of these spectra, there is a degeneracy between temperature, gravity,
and metallicity. For most stars, assuming a lower metallicity results
in a spectrum that looks hotter and/or has a weaker surface gravity.
Thus for the initial characterization we assume solar metallicity.
This allows us to determine the astrophysical parameters with a
typical precision of 150 K for effective temperature, 1 dex for log
surface gravity, and 1 \kms\ for the projected rotational velocity,
although the rotation becomes more sensitive to systematic errors for
stars rotating more slowly than the instrumental resolution of 8.5
\kms, and is not reliable for stars rotating more rapidly than about
120 \kms.  The rotation is relatively insensitive to the temperature,
gravity, and metallicity.  However, the temperature and gravity may
have systematic errors that are several times larger than the
precisions quoted above if the metallicity is substantially different
from solar.  As a rough rule of thumb, for solar-type stars a decrease
in the metallicity of 0.5 dex corresponds to an increase in the
temperature of about 200 K.

The strategy of most wide-angle ground-based photometric surveys for
transiting planets, including HATNet, is to analyze every point source
down to some limiting magnitude whose value is set by the desired
photometric performance.  Thus many of the HATNet candidates prove to
be evolved.  If the surface gravity that we derive from the CfA
spectra is weaker than $\log g=3.25$ and there is no significant
rotation, we assume that either the HATNet detection is a photometric
false alarm, or that the primary star is a giant diluting the light
of an eclipsing binary that is unresolved in the HATNet images, either
a physically bound system or a chance alignment.  Normally we put
aside the targets classified as giants and do not invest additional
resources in follow-up observations.

If the CfA spectra indicate that a candidate has a temperature and
gravity consistent with a star on or near the Main Sequence and the
initial spectra show no velocity variations, we usually schedule
additional observations the next month to make sure the velocity does
not vary. If the candidate survives this stage, only then do we try to
schedule photometric observations with KeplerCam, with the goal of
determining a high-quality light curve for a transit event.  However,
this step is not practical if the photometric ephemeris is not
accurate enough to predict the time of transit within a few hours.
The failure to confirm a transit with high-quality photometry is
ambiguous, and can not distinguish a photometric false alarm from
an inadequate ephemeris.  Confirmation that a candidate undergoes
transits is an area where smaller telescopes can contribute, such as
sophisticated facilities operated by amateurs \citep{McCullough:06} or
the network of photometric telescopes planned for the Las Cumbres
Observatory Global Telescope Network \citep{Brown:07}.

Because the HATNet pixels are so large on the sky (about $14\arcsec$
for the original 2Kx2K CCDs used to observe field G205 and $9\arcsec$
for the 4Kx4K CCDs now in use) many of the candidates prove to
have close companions when viewed with much higher spatial resolution.
In some cases this allows us to use KeplerCam to demonstrate that the
transit-like light curve is actually due to a nearby eclipsing binary
diluted by the brighter target star.

Candidates that show no velocity variation at the level of a few
hundred \ms\ with the reconnaissance spectroscopy and have a
good-quality light curve consistent with a transiting planet are then
followed up for very precise velocities with facilities such as HIRES
on Keck I, HRS on Subaru, and/or Sophie at OHP, in order to derive
orbital solutions for the host star and thus the mass of the planet
relative to the host star.

\subsection{Reconnaissance Spectroscopy}

Starting in January 2004, all but 4 of the 32 transiting-planet candidates
identified in HATNet field G205 were observed with the CfA Digital 
Speedometers (Latham 1992).  The remaining 4 are faint and have not yet been
observed.  Of those observed, all but one yielded reliable radial
velocities, as described below. The only one for which we were unable
to determine a velocity was HTR205-014, whose spectrum proved to be
featureless in the Mg b region.  In retrospect, this result is not
surprising, given this star's very hot 2MASS color, $J-K_{\rm s} = 0.13$
mag.  In the early days, when we were anxious to confirm our first
HATNet transiting planet, we followed up everything that looked like
it might be a planet, no matter what its color or proper motion.
However, 2MASS is an extremely valuable resource for wide-angle
surveys.  The 2MASS colors from the near infrared are relatively
insensitive to reddening and thus provide a good initial guess for the
temperature of a star.  Moreover, the 2MASS spatial resolution is much
better than the HATNet cameras, and 2MASS can be used to track
down nearby companions that contaminate the HATNet images.  Finally,
the 2MASS astrometry is of reliably high quality and is quite useful
in the reduction and analysis of images.

The radial velocities of the 27 candidates in field G205 with reliable
measurements are reported for all the individual observations in
Tables~\ref{tab:slvel} and \ref{tab:dlvel}. The velocity results for
the 19 stars in Table~\ref{tab:slvel} are based on one-dimensional
correlations using {\bf rvsao} running inside IRAF\footnote{IRAF (Image
Reduction and Analysis Facility) is distributed by the National
Optical Astronomy Observatories, which are operated by the Association
of Universities for Research in Astronomy, Inc., under contract with
the National Science Foundation.} and templates from
our library of synthetic spectra calculated by Jon Morse \citep[e.g.,
see][]{Latham:02}.  For four of the targets (HTR205-004, HTR205-010,
HTR205-011, and HTR205-020) the double-lined nature of the spectra
was obvious from the one-dimensional correlation analysis of the first
two exposures, and the velocities for both components of these
binaries are listed in Table~\ref{tab:dlvel} based on two-dimensional
correlations using TODCOR \citep{Zucker:94} as implemented at CfA by
GT\@.  Eventually it was possible to use TODCOR to dig out secondary
velocities for two additional binaries (HTR205-012 and HTR205-017),
once single-lined orbital solutions were available and could be used
to guide the search for the secondary correlation peaks.  The
parameters of the template spectra that were used to derive the
velocities are reported in Tables~\ref{tab:slvel} and \ref{tab:dlvel},
encoded in the form tTTTTTgGGp00vVVV, where TTTTT gives the effective
temperature in K, GG the log surface gravity times 10 in cgs units,
p00 signifies that the metallicity [m/H] is plus 0, i.e.~solar, and
VVV is the rotational velocity in \kms.  In the case of the spectra
analyzed with TODCOR, the choice of primary template was based on
grids of one-dimensional correlations, while the choice of secondary
template used the results of the orbital solution for guidance.

\ifthenelse{\boolean{emulateapj}}{
	\begin{deluxetable*}{llllrr}
}{
	\begin{deluxetable}{llllrr}	
}
\tabletypesize{\scriptsize}
\tablewidth{0pc}
\tablenum{2}
\tablecaption{Single-Lined Radial Velocities}
\label{tab:slvel}
\tablehead{
\colhead{~~~~~Star~~~~~~~~~~} &
\colhead{Tel} &
\colhead{Template} &
\colhead{HJD} &
\colhead{$V_{\rm A}$} &
\colhead{$\sigma(V_{\rm A})$}}
\startdata
HTR205-001 & T & t10000g45p00v008 & 2453158.90513 &   -2.58 &  0.65 \\
HTR205-001 & T & t10000g45p00v008 & 2453161.96505 &   -2.75 &  0.58 \\
HTR205-001 & W & t10000g45p00v008 & 2453183.82621 &   -2.19 &  1.17 \\
HTR205-002 & W & t05000g30p00v000 & 2453137.83494 &  -27.03 &  0.27 \\
HTR205-002 & W & t05000g30p00v000 & 2453154.80255 &  -27.06 &  0.28 \\
HTR205-002 & W & t05000g30p00v000 & 2453180.82534 &  -26.76 &  0.26 \\
HTR205-003 & T & t06500g45p00v030 & 2453162.91121 &  -23.55 &  1.27 \\
HTR205-003 & T & t06500g45p00v030 & 2453183.95068 &  -67.55 &  1.48 \\
HTR205-003 & T & t06500g45p00v030 & 2453511.95989 &   -1.53 &  2.43
\enddata
\tablecomments{
column (1): HATNet transiting-planet candidate designation, \\
column (2): telescope (W=Wyeth, T=Tillinghast), \\
column (3): template (see text for code), \\
column (4): heliocentric Julian date, \\
column (5): heliocentric radial velocity (\kms), \\
column (6): radial-velocity error estimate (\kms).\\
This table is presented in its entirety in the electronic edition of the Astrophysical Journal.
A portion is shown here for guidance regarding its form and content.}
\ifthenelse{\boolean{emulateapj}}{
	\end{deluxetable*}
}{
	\end{deluxetable}
}

\ifthenelse{\boolean{emulateapj}}{
	\begin{deluxetable*}{llllrr}
}{
	\begin{deluxetable}{llllrr}	
}
\tabletypesize{\scriptsize}
\tablewidth{0pc}
\tablenum{3}
\tablecaption{Double-Lined Radial Velocities}
\label{tab:dlvel}
\tablehead{
\colhead{~~~~~Star~~~~~~~~~~} &
\colhead{Template A}     &
\colhead{Template B}     &
\colhead{HJD}            &
\colhead{$V_{\rm A}$}    &
\colhead{$V_{\rm B}$}}
\startdata
HTR205-004 & t06250g40p00v016 & t06250g40p00v016 & 24453162.9343 &  -41.09 &  -16.87 \\
HTR205-004 & t06250g40p00v016 & t06250g40p00v016 & 24453183.9643 &   22.52 &  -83.60 \\
HTR205-004 & t06250g40p00v016 & t06250g40p00v016 & 24453188.9069 &   45.65 & -102.90 \\
HTR205-004 & t06250g40p00v016 & t06250g40p00v016 & 24453684.6943 &   37.57 & -102.34 \\
HTR205-004 & t06250g40p00v016 & t06250g40p00v016 & 24453689.6640 &    5.36 &  -72.24 \\
HTR205-004 & t06250g40p00v016 & t06250g40p00v016 & 24453694.6542 &  -23.91 &  -36.67 \\
HTR205-004 & t06250g40p00v016 & t06250g40p00v016 & 24453714.6282 &  -97.31 &   39.29 \\
HTR205-004 & t06250g40p00v016 & t06250g40p00v016 & 24453746.5850 &  -41.46 &  -19.67 \\
HTR205-004 & t06250g40p00v016 & t06250g40p00v016 & 24453984.8431 &   42.79 & -102.59
\enddata
\tablecomments{
column (1): HATNet transiting-planet candidate designation, \\
column (3): template for primary, \\
column (3): template for secondary, \\
column (4): heliocentric Julian date, \\
column (5): primary heliocentric radial velocity (\kms), \\
column (6): secondary heliocentric radial velocity (\kms). \\
This table is presented in its entirety in the electronic edition of
the Astrophysical Journal. A portion is shown here for guidance
regarding its form and content.}
\ifthenelse{\boolean{emulateapj}}{
	\end{deluxetable*}
}{
	\end{deluxetable}
}

The mean velocities and various indicators of the uncertainties are
presented in Table~\ref{tab:sumvel}.  For those binaries with orbital
solutions, the mean velocity and error are actually the center-of-mass
velocity and error estimate from the orbital solution.  Otherwise the
error of the mean velocity is one of the following two, whichever is
larger: the actual observed rms velocity residuals divided by the
square root of the number of observations, or the average of the
internal velocity error estimates, again divided by root $N_{\rm
obs}$.  $P(\chi^2)$ is the probability of observing velocity residuals
and corresponding $\chi^2$ larger than the actual observed residuals
and $\chi^2$, assuming a star with constant velocity and Gaussian
errors for the observed velocities.  Experience shows that stars with
$P(\chi^2)$ values less than 0.001 usually prove to be binaries.  For
10 of the candidates the radial-velocity data were sufficient to allow
orbital solutions.  The orbital parameters are reported in
Tables~\ref{tab:ssborb} and \ref{tab:dsborb}.

\ifthenelse{\boolean{emulateapj}}{
	\begin{deluxetable*}{lrrrrrrrrrrrl}
}{
	\begin{deluxetable}{lrrrrrrrrrrrl}
}
\tabletypesize{\scriptsize}
\tablewidth{0pc}
\tablenum{4}
\tablecaption{Mean Radial Velocities and Error Estimates}
\label{tab:sumvel}
\tablehead{
\colhead{~~~~~~~Star~~~~~~~~~~~} &
\colhead{$N_{\rm obs}$} &
\colhead{Span} &
\colhead{$\langle V_{\rm rot} \rangle$} &
\colhead{$\langle V_{\rm rad} \rangle$} &
\colhead{$\pm$} &
\colhead{ext} &
\colhead{int} &
\colhead{e/i} &
\colhead{$\chi^2$} &
\colhead{P($\chi^2$)} &
\colhead{$\langle$ht$\rangle$} &
\colhead{Comments}}
\startdata
HTR205-001 \df &  3 &  24 &  8.0 & $ -2.48$ & 0.48 & 0.36 & 0.83 &  0.43 &   0.35 & 0.838 & 0.64 & P?    \\
HTR205-002 \df &  3 &  43 &  0.5 & $-26.96$ & 0.21 & 0.16 & 0.37 &  0.44 &   0.40 & 0.820 & 0.93 & G,WD  \\
HTR205-003 \df & 17 & 879 & 31.0 & $-34.96$ & 0.49 & 1.77 & 1.88 &  0.94 &   \nd  & \nd   & 0.68 & EB,SO \\
HTR205-004 \df & 12 & 882 & 17.3 & $-30.92$ & 0.53 & 3.06 & 2.10 &  1.45 &   \nd  & \nd   & \nd  & EB,DO \\
HTR205-005 \df &  2 &  25 & 60.2 & $-16.17$ & 26.8 & 37.9 & 6.18 &  6.14 &  52.48 & 0.000 & 0.21 & S,FR,WD \\
HTR205-006 \df &  2 & 504 &  0.0 & $-51.40$ & 0.37 & 0.53 & 0.38 &  1.39 &   1.93 & 0.165 & 0.91 & G,WD  \\
HTR205-007W \df & 4 & 793 & 53.0 & $-17.44$ &15.32 &30.63 & 2.67 & 11.45 & 508.86 & 0.000 & 0.54 & S,FR  \\
HTR205-007E \df & 2 &   2 &  7.4 & $-36.25$ & 0.73 & 1.04 & 0.80 &  1.29 &   1.68 & 0.712 & 0.71 & P?    \\
HTR205-008 \df &  3 &  22 &  3.0 & $+31.65$ & 0.22 & 0.22 & 0.39 &  0.57 &   0.62 & 0.733 & 0.91 & G,WD  \\
HTR205-009 \df &  1 &   0 &  120 & $ +2.50$ & 5.16 & \nd  & 5.16 &  \nd  &   \nd  & \nd   & 0.34 & FR    \\
HTR205-010 \df & 11 & 362 & 45.8 & $-11.67$ & 0.78 & 2.44 & 3.83 &  0.64 &   \nd  & \nd   & 0.56 & EB,DO \\
HTR205-011 \df & 11 & 804 & 22.1 & $ -6.22$ & 0.39 & 2.01 & 4.43 &  0.45 &   \nd  & \nd   & 0.64 & EB,DO \\
HTR205-012 \df & 13 & 832 & 36.0 & $-30.75$ & 0.73 & 3.16 & 2.75 &  1.15 &   \nd  & \nd   & 0.57 & EB,DO \\
HTR205-013 \df & 23 & 630 & 28.9 & $ -9.74$ & 0.23 & 1.01 & 1.24 &  0.82 &   \nd  & \nd   & 0.80 & EB,SO \\
HTR205-014 \df &  2 &  31 &  \nd &    \nd   & \nd  & \nd  & \nd  &  \nd  &   \nd  & \nd   & \nd  & FR    \\
HTR205-015 \df &  2 & 386 &  3.0 & $-42.64$ & 0.35 & 0.26 & 0.50 &  0.53 &   0.30 & 0.587 & 0.88 & G     \\
HTR205-016 \df &  3 & 504 &  0.0 & $-30.05$ & 0.25 & 0.22 & 0.44 &  0.49 &   0.53 & 0.765 & 0.88 & G,WD  \\
HTR205-017 \df & 12 & 804 & 10.6 & $ -9.03$ & 0.31 & 0.94 & 0.76 &  1.23 &   \nd  & \nd   & 0.73 & EB,DO \\
HTR205-018 \df &  5 & 114 &  5.9 & $-23.07$ & 0.23 & 0.28 & 0.52 &  0.54 &   1.50 & 0.826 & 0.86 & NEB   \\
HTR205-019 \df & 10 & 862 &  0.0 & $-38.85$ & 0.20 & 0.31 & 0.44 &  0.71 &   0.71 & \nd   & 0.89 & G,SO,WD \\
HTR205-020 \df & 10 & 804 & 13.1 & $-12.26$ & 0.30 & 0.97 & 1.33 &  0.73 &   \nd  & \nd   & 0.49 & EB,DO \\
HTR205-021 \df & 12 & 919 &  0.5 & $ -6.86$ & 0.11 & 0.21 & 0.48 &  0.44 &   \nd  & \nd   & 0.88 & SO    \\
HTR205-022 \df &  6 & 233 &  7.5 & $ -3.44$ & 0.19 & 0.32 & 0.47 &  0.68 &   2.02 & 0.846 & 0.88 & NP    \\
HTR205-023 \df &  6 & 233 &  1.5 & $ -2.94$ & 0.23 & 0.55 & 0.49 &  1.14 &   6.19 & 0.288 & 0.88 & P     \\
HTR205-024 \df &  8 & 357 & 12.8 & $-22.53$ & 0.28 & 0.79 & 0.64 &  1.23 &  10.23 & 0.176 & 0.85 & P     \\
HTR205-025 \df &  2 &   4 &  0.8 & $-42.63$ & 0.27 & 0.27 & 0.38 &  0.71 &   0.50 & 0.478 & 0.92 & G,WD  \\
HTR205-026 \df &  4 & 293 &  0.5 & $-17.49$ & 0.26 & 0.34 & 0.51 &  0.66 &   1.09 & 0.780 & 0.82 & WD    \\
HTR205-027 \df &  0 \\
HTR205-028 \df &  0 \\
HTR205-029 \df &  0 \\
HTR205-030 \df &  4 & 214 &  3.4 & $-23.67$ & 0.45 & 0.90 & 0.59 &  1.52 &   7.44 & 0.059 & 0.823 & WD   \\
HTR205-031 \df &  0
\enddata
\tablecomments{
column (1): HATNet transiting-planet candidate designation, \\
columns (2 and 3): number of observations and time spanned (days), \\
column (4): projected rotational velocity for the primary derived from
one-dimensional correlations (\kms), \\
column (5): mean radial velocity or center-of-mass velocity from an
orbital solution (\kms), \\
column (6): error in the mean velocity (\kms), \\
column (7): external rms residuals in the observed velocities of the
primary; from the orbital fit in the case of a binary (\kms), \\
column (8): mean internal velocity error estimate (\kms), \\
column (9): ratio of external to internal errors, \\
columns (10 and 11): $\chi^2$ and $\chi^2$ probability, \\
column (12): mean of the peak correlation height, \\
column (13): Comments: DO=double-lined
orbital solution, EB=eclipsing binary, FR=rapidly rotating star,
G=giant, NEB=blend with nearby eclipsing binary,
NP=blend with nearby transiting planet,
S=velocity is variable, SO=single-lined orbital solution,
P=planet confirmed by a solution for the spectroscopic orbit,
P?=possible transiting planet,WD=withdrawn by HATNet as a photometric false alarm.}
\ifthenelse{\boolean{emulateapj}}{
	\end{deluxetable*}
}{
	\end{deluxetable}
}

\subsubsection{Spectroscopic Binaries with Orbital Solutions}

\begin{figure}[!ht]
\plotone{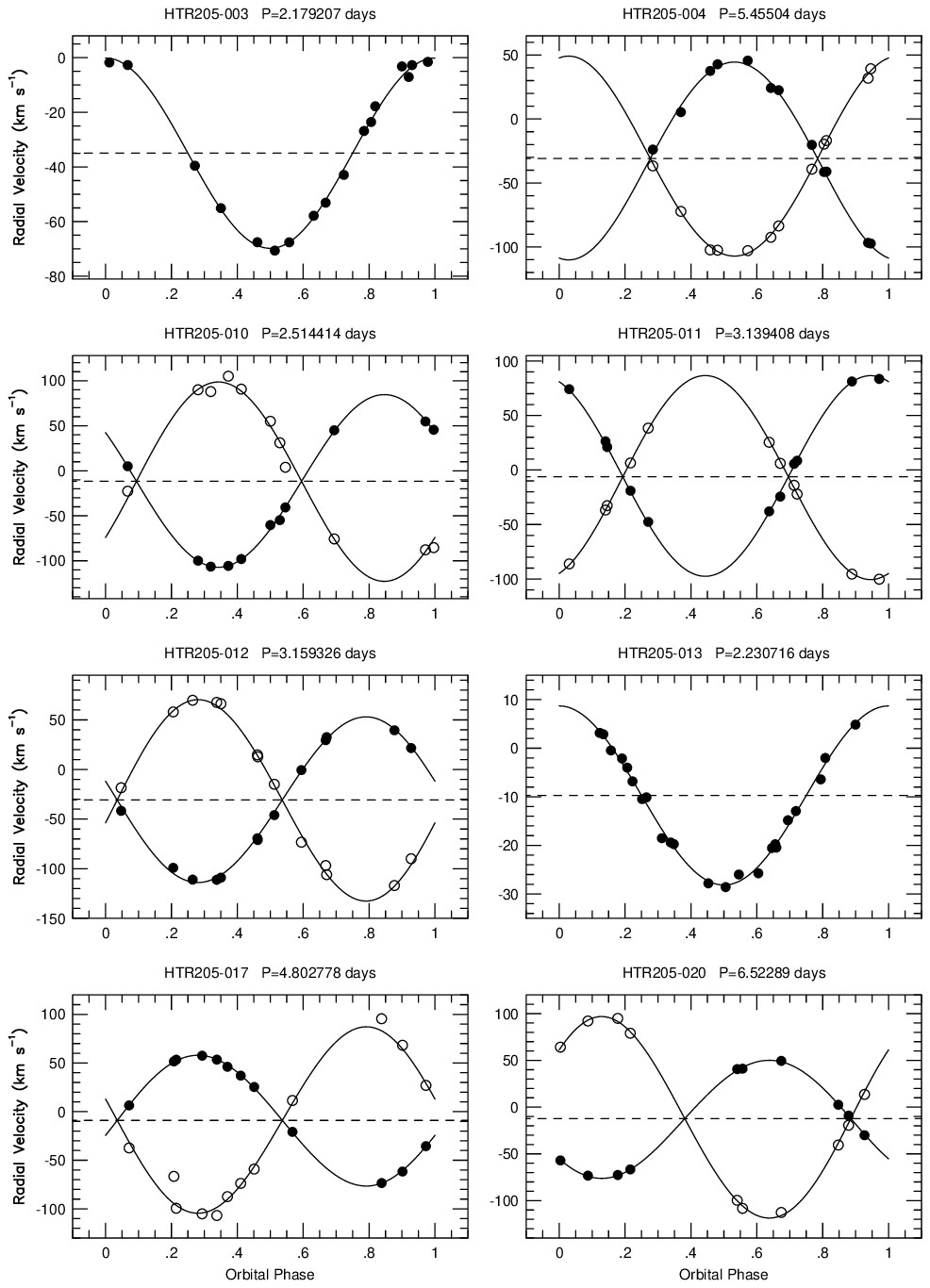}
\caption{
The velocity curves for the eight confirmed eclipsing binaries, phased
to the orbital solution reported in Tables~\ref{tab:ssborb} and
\ref{tab:dsborb}, together with the observed velocities.}
\label{fig:eborb}
\end{figure}

Ten of the candidates with variable velocities yielded orbital
solutions.  For eight of these (see Figure~\ref{fig:eborb}:
HTR205-003, HTR205-004, HTR205-010, HTR205-011, HTR205-012,
HTR205-013, HTR205-017, and HTR205-020), the orbital period is
consistent with the photometric period from HATNet within the errors.
This agreement proves beyond a reasonable doubt that the observed
light curves are actually due to eclipses by stellar companions. Note
that for six of these binaries the orbital period is actually twice
the original HATNet photometric period.  When the primary and secondary
eclipses are not very different and the orbit is circular, the
analysis of the photometry for periodic events often settles on half
the true period.  This implies that the eclipses must be grazing, in
order to produce the small observed dips.  The two exceptions are the
eclipsing binaries HTR205-003 and HTR205-013, where the companions are M dwarfs and the
secondary eclipse is much too weak for detection with HATNet, despite
the fact that the eclipses are full.  A detailed analysis of
HTR205-013 has already been published by \citet{Beatty:07}.

%

\ifthenelse{\boolean{emulateapj}}{
	\begin{deluxetable*}{lcrrrrrrrrr}
}{
	\begin{deluxetable}{lcrrrrrrrrr}
}
\tabletypesize{\scriptsize}
\tablenum{5}
\tablewidth{0pt}
\tablecaption{Single-Lined Orbital Solutions}
\label{tab:ssborb}
\tablehead{
\colhead{~~~~~Star~~~~~~~~~~} &
\colhead{$P$} &
\colhead{$\gamma$} &
\colhead{$K$} &
\colhead{$e$} &
\colhead{$\omega$} &
\colhead{$T$} &
\colhead{$a_{\rm A} \sin i$} &
\colhead{$f(m)$} &
\colhead{$N$} &
\colhead{Span}\\
\colhead{} &
\colhead{} &
\colhead{} &
\colhead{} &
\colhead{} &
\colhead{} &
\colhead{} &
\colhead{} &
\colhead{} &
\colhead{$\sigma$} &
\colhead{Cycles}}
\startdata
HTR205-003 \df &   2.178207 &$ -34.96 $&  34.84 &    0.0  &  \nd   &  53599.1750  &   1.044 &   0.0096  &   17 & 879.8 \\
               &\p 0.000069 &\p  0.49  &\p 0.60 &    \nd  &  \nd   &\p    0.0071  &\p 0.075 &\p 0.0021  & 1.77 & 403.7 \\
HTR205-013 \df &   2.230716 &$  -9.74 $&  18.42 &    0.0  &  \nd   &  53197.72708 &   0.565 &   0.00144 &   23 & 630.2 \\
               &\p 0.000049 &\p  0.23  &\p 0.31 &    \nd  &  \nd   &\p    0.00022 &\p 0.029 &\p 0.00022 & 1.01 & 282.5 \\
HTR205-019 \df &  94.331    &$ -38.85 $&  25.48 &   0.408 &  289.1 &  53738.53    &  30.17  &   0.1230  &   10 & 862.7 \\
               &\p 0.035    &\p  0.20  &\p 0.33 &\p 0.010 &\p  1.4 &\p    0.21    &\p 0.18  &\p 0.0022  & 0.31 &   9.1 \\
HTR205-021 \df & 113.301    &$  -6.86 $&  20.85 &   0.364 &  238.2 &  53653.53    &  30.248 &   0.08591 &   12 & 919.6 \\
               &\p 0.056    &\p  0.011 &\p 0.12 &\p 0.006 &\p  1.1 &\p    0.23    &\p 0.058 &\p 0.00052 & 0.21 &   8.1
\enddata
\tablecomments{
column (1): HATNet transiting-planet candidate designation, \\
column (2): period $P$ in days, \\
column (3): center-of-mass velocity $\gamma$ in \kms, \\
column (4): projected orbital semi-amplitude of the primary $K$ in \kms,\\
column (5): eccentricity $e$, assumed circular if no error \\
column (6): angle of periastron $\omega$ in degrees, \\
column (7): heliocentric Julian Date $-$2,400,000 for periastron passage $T$, or for maximum velocity if $e=0$ assumed, \\
column (8): projected semi-major axes of the primary $a \sin i$ in Gm, \\
column (9): mass function $f(m)$ in \msun, \\
column (10): number of velocities and rms velocity residuals in \kms, \\
column (11): time spanned by the observations in days and number of orbital cycles covered.}
\ifthenelse{\boolean{emulateapj}}{
	\end{deluxetable*}
}{
	\end{deluxetable}
}

\ifthenelse{\boolean{emulateapj}}{
	\begin{deluxetable*}{lcrrrrrrr}
}{
	\begin{deluxetable}{lcrrrrrrr}
}
\tabletypesize{\scriptsize}
\tablenum{6}
\tablewidth{0pt}
\tablecaption{Double-Lined Orbital Solutions}
\label{tab:dsborb}
\tablehead{
\colhead{~~~~~Star~~~~~~~~~~} &
\colhead{$P$} &
\colhead{$\gamma$} &
\colhead{$K_{\rm A}$} &
\colhead{$T$} &
\colhead{$a_{\rm A} \sin i$} &
\colhead{$M_{\rm A} \sin^3 i$} &
\colhead{$N$} &
\colhead{Span}\\
\colhead{} &
\colhead{$q$} &
\colhead{} &
\colhead{$K_{\rm B}$} &
\colhead{} &
\colhead{$a_{\rm B} \sin i$} &
\colhead{$M_{\rm B} \sin^3 i$} &
\colhead{$\sigma$} &
\colhead{Cycles}}
\startdata
HTR205-004 \df &   5.45504  &$  -30.92 $&  77.3  &  53633.0976 &   5.80  &   1.067 &    12 & 882.8 \\
               &\p 0.00014  &\p   0.53  &\p 1.4  &\p    0.0077 &\p 0.12  &\p 0.035 &   3.1 & 181.8 \\
               &     0.989  &\nd        &  78.2  &\nd          &   5.86  &   1.056 &    12 &\nd    \\
               &\p   0.024  &\nd        &\p 1.0  &\nd          &\p 0.08  &\p 0.044 &   2.0 &\nd    \\
HTR205-010 \df &   2.514414 &$  -11.67 $&  95.9  &  53876.8713 &   3.316 &   1.23  &    11 & 362.9 \\
               &\p 0.000082 &\p   0.78  &\p 1.2  &\p    0.0048 &\p 0.048 &\p 0.10  &   2.4 & 144.3 \\
               &     0.866  &\nd        & 110.7  &\nd          &   3.83  &   1.067 &    11 &\nd    \\
               &\p   0.036  &\nd        &\p 3.8  &\nd          &\p 0.15  &\p 0.054 &   8.8 &\nd    \\
HTR205-011 \df &   3.139408 &$   -6.22 $&  92.0  &  53715.7397 &   3.973 &   1.049 &    11 & 804.9 \\
               &\p 0.000047 &\p   0.39  &\p 1.2  &\p    0.0026 &\p 0.057 &\p 0.026 &   2.0 & 256.4 \\
               &     0.983  &\nd        &  93.6  &\nd          &   4.041  &  1.032 &    11 &\nd    \\
               &\p   0.018  &\nd        &\p 0.9  &\nd          &\p 0.043 &\p 0.031 &   1.5 &\nd    \\
HTR205-012 \df &   3.159326 &$  -30.75 $&  83.5  &  53683.3584 &   3.627 &   1.137 &    13 & 832.8 \\
               &\p 0.000058 &\p   0.73  &\p 1.3  &\p    0.0056 &\p 0.064 &\p 0.052 &   3.2 & 263.6 \\
               &     0.823  &\nd        & 101.5  &\nd          &   4.409 &   0.935 &    13 &\nd    \\
               &\p   0.021  &\nd        &\p 1.8  &\nd          &\p 0.089 &\p 0.038 &   4.4 &\nd    \\
HTR205-017 \df &   4.802778 &$   -9.03 $&  67.25 &  53678.3002 &   4.441 &   1.27  &    12 & 803.9 \\
               &\p 0.000096 &\p   0.31  &\p 0.43 &\p    0.0051 &\p 0.032 &\p 0.14  &   0.9 & 167.4 \\
               &     0.701  &\nd        &  95.9  &\nd          &   6.33  &   0.890 &    12 &\nd    \\
               &\p   0.036  &\nd        &\p 4.2  &\nd          &\p 0.31  &\p 0.052 &  10.8 &\nd    \\
HTR205-020 \df &   6.52289  &$  -12.26 $&  63.16 &  53703.7368 &   5.664 &   2.128 &    10 & 804.0 \\
               &\p 0.00015  &\p   0.30  &\p 0.30 &\p    0.0065 &\p 0.043 &\p 0.059 &   1.0 & 123.3 \\
               &     0.586  &\nd        & 107.7  &\nd          &   9.67  &   1.247 &    10 &\nd    \\
               &\p   0.008  &\nd        &\p 1.1  &\nd          &\p 0.12  &\p 0.025 &   2.7 &\nd
\enddata
\tablecomments{
column (1): HATNet transiting-planet candidate designation, \\
column (2): period $P$ in days, \\
column (3): center-of-mass velocity $\gamma$ in \kms,\\
column (4): projected orbital velocities of the primary and secondary $K_{\rm A}$ and $K_{\rm B}$ in \kms,\\
column (5): heliocentric Julian Date $-$2,400,000 for time of maximum velocity $T$,\\
column (6): projected semi-major axes of the primary and secondary $a_{\rm A} \sin i$ and $a_{\rm B} \sin i$ in Gm,\\
column (7): projected masses of the primary and secondary $M_{\rm A} \sin^3 i$ and $M_{\rm B} \sin^3 i$ in \msun,\\
column (8): number of velocities and rms velocity residuals in \kms\ for the primary and secondary, and\\
column (9): time spanned by the observations in days and number of orbital cycles covered.}
\ifthenelse{\boolean{emulateapj}}{
	\end{deluxetable*}
}{
	\end{deluxetable}
}

For all eight of the confirmed eclipsing binaries the orbital
solutions yield eccentricities which are indistinguishable from zero
when the eccentricity is allowed to be a free parameter in the
solution (see Table~\ref{tab:eccorb}), presumably because the orbits
have been circularized by tidal mechanisms.  Thus for the final
orbital solutions we adopted $e\equiv0$, and the epochs for these
binaries listed in Tables~\ref{tab:ssborb} and \ref{tab:dsborb} are
the times of maximum velocity rather than the times of periastron
passage.

\ifthenelse{\boolean{emulateapj}}{
	\begin{deluxetable}{lll}
}{
	\begin{deluxetable}{lll}
}
\tabletypesize{\scriptsize}
\tablenum{7}
\tablewidth{0pt}
\tablecaption{Orbital Eccentricities}
\label{tab:eccorb}
\tablehead{
\colhead{~~~~~Star~~~~~~~~~~} &
\colhead{$e$}                 &
\colhead{$\sigma(e)$}}
\startdata
HTR205-003 & 0.025  &\p 0.020  \\
HTR205-004 & 0.025  &\p 0.011  \\
HTR205-010 & 0.007  &\p 0.016  \\
HTR205-011 & 0.0085 &\p 0.0065 \\
HTR205-012 & 0.000  &\p 0.011  \\
HTR205-013 & 0.012  &\p 0.021  \\
HTR205-017 & 0.0318 &\p 0.0055 \\
HTR205-019 & 0.408  &\p 0.010  \\
HTR205-020 & 0.0191 &\p 0.0074 \\
HTR205-021 & 0.364  &\p 0.006
\enddata
\tablecomments{
column (1): HATNet transiting-planet candidate designation, \\
column (2): orbital eccentricity $e$ when allowed to be a free parameter. \\
column (3): estimated standard deviation of $e$.}
\ifthenelse{\boolean{emulateapj}}{
	\end{deluxetable}
}{
	\end{deluxetable}
}

The preliminary orbital solutions presented here and the HATNet
discovery light curves are not good enough to justify full-blown
analyses of the confirmed eclipsing binaries.  However, some of these
systems may merit further investigation.  To provide some guidance for
such efforts, we summarize some of the expected characteristics of
these binaries in Table~\ref{tab:ebsum}.  As an estimate of the
mass of each primary in a double-lined system, we list the minimum
mass from the orbital solution.  This is the same as the actual mass
if the orbit is viewed exactly edge-on.  To estimate the radius of the
primary, we invoke the broadening of the lines that we derive from the
CfA spectra. Tidal mechanisms tend to align the stellar rotation axes
with the orbital axis, to synchronize the rotational periods with the
orbital period, and to circularize the orbit, normally with the
sequence of events in this order \citep[e.g.\ see][]{Zahn:89}.
Because all of our eclipsing binaries show circular orbits, we can
reasonably assume that the rotation of the stars has been synchronized
and aligned with the orbit, and thus we can use the projected
rotational velocity, $v \sin i_{\rm rot}$, to estimate the radius of
the primary, $R_{\rm A}$:
\begin{eqnarray}
R_{\rm A} = 0.0198 (v \sin i_{\rm rot} P_{\rm rot}) / \sin i_{\rm rot},
\end{eqnarray}
where $P_{\rm rot} \equiv P_{\rm orb}$ and $i_{\rm rot} \equiv i_{\rm
orb}$ by assumption, and the coefficient 0.0198 delivers the
radius in solar units if the rotational broadening is in \kms\ and the
period is in days.

\ifthenelse{\boolean{emulateapj}}{
	\begin{deluxetable*}{lrrrrrrrr}
}{
	\begin{deluxetable}{lrrrrrrrr}
}
\tabletypesize{\scriptsize}
\tablenum{8}
\tablewidth{0pt}
\tablecaption{Eclipsing Binary Radii}
\label{tab:ebsum}
\tablehead{
\colhead{~~~~~Star~~~~~~~~~~}   &
\colhead{$v \sin i_{\rm rot}$}  &
\colhead{$P_{\rm rot}$}         &
\colhead{$R_{\rm A}$}           &
\colhead{$M_{\rm A} \sin^3 i$}  &
\colhead{$q$}                   &
\colhead{$T_{\rm eff}$}         &
\colhead{$\log g$}              &
\colhead{$L_{\rm B}/L_{\rm A}$}
}
\startdata
HTR205-003 & 31.0  & 2.179 & 1.34 & \nd  & \nd  & 6500 & 4.5 & \nd   \\
HTR205-004 & 17.3  & 5.455 & 1.87 & 1.07 & 0.99 & 6250 & 4.0 & 1.00  \\
HTR205-010 & 45.8  & 2.514 & 2.28 & 1.23 & 0.87 & 5750 & 3.5 & 0.36  \\
HTR205-011 & 22.1  & 3.139 & 1.37 & 1.05 & 0.98 & 5500 & 3.5 & 0.90  \\
HTR205-012 & 36.0  & 3.159 & 2.25 & 1.13 & 0.82 & 6250 & 4.0 & 0.24  \\
HTR205-013 & 28.9  & 2.231 & 1.27 & \nd  & \nd  & 6250 & 4.0 & \nd   \\
HTR205-017 & 10.6  & 4.803 & 1.01 & 1.26 & 0.70 & 6750 & 4.0 & 0.09  \\
HTR205-020 & 13.1  & 6.523 & 1.69 & 2.12 & 0.59 &11500 & 5.0 & 0.09
\enddata
\tablecomments{
column (1): HATNet transiting-planet candidate designation, \\
column (2): spectroscopic projected rotational velocity $v \sin i_{\rm rot}$ in \kms,
            where the inclination of the rotational axis, $i_{\rm rot}$ is assumed to
            be the same as the inclination of the orbital axis, $i_{\rm orb}$, \\
column (3): rotational period, $P_{\rm rot}$, assumed equal to the orbital period $P$, \\
column (4): radius of the primary, $R_{\rm A}$, assuming synchronized and aligned rotation, \\
column (5): minimum mass of the primary in a double-lined binary, $M_{\rm A} \sin^3 i$, \\
column (6): secondary to primary mass ratio, $M_{\rm B}/M_{\rm A}$, \\
column (5): spectroscopic effective temperature estimate for the primary, $T_{\rm eff}$, \\
column (6): spectroscopic log surface gravity estimate for the primary, $\log g$, \\
column (7): luminosity ratio at 5187 \AA\ from TODCOR, $L_{\rm B}/L_{\rm A}$.}
\ifthenelse{\boolean{emulateapj}}{
	\end{deluxetable*}
}{
	\end{deluxetable}
}

\begin{figure}[!h]
\includegraphics[bb=60 560 550 750, clip=true, width=88mm]{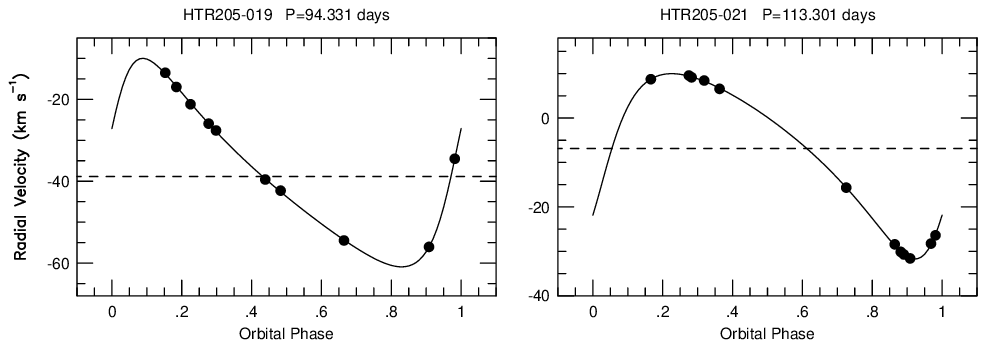}
\caption{
The velocity curves and observed velocities for the two
single-lined binaries with orbital solutions reported in
Table~\ref{tab:ssborb} that do not match the photometric ephemerides.
In particular, the orbital periods are considerably longer than the
photometric periods.
\label{fig:noneborb}}
\end{figure}

For two of the ten spectroscopic binaries with orbital solutions
(plotted in Figure~\ref{fig:noneborb}: HTR205-019 and HTR205-021), the
orbital period is much longer than the photometric period from HATNet.
Thus the stellar companions responsible for the orbital motion can not
be responsible for the observed dips in the HATNet light curves for
these candidates.  A possible explanation is that the light of the
observed primary is diluting the light from a nearby eclipsing binary
that is not spatially resolved by HATNet, either in a physical system
or a chance alignment. If the eclipsing binary is bound to the
primary, then the orbital motion observed for the primary may reflect
its orbit around the eclipsing binary, in which case only three stars
are needed to explain the observations.  If the separation between
the primary and the eclipsing binary is very wide, either in a
long-period orbit or because the alignment is due to chance, then four
stars are required to explain the observations. Support for the
interpretation that an observed HATNet light curve arises from a
diluted eclipsing binary can come from a better light curve observed
with a larger telescope.  Higher spatial resolution may reveal that a
nearby eclipsing binary was blended with the target in the HATNet
images and is the source of the transit-like event.  Even if the
eclipsing binary is not resolved spatially, better photometric
precision may be able to show that the ingress and egress last longer
than allowed by the observed eclipse depth, thus indicating that the
eclipsing objects are larger than implied by the depth because of
dilution by the primary star.  Another possible explanation is that
the HATNet detection is a photometric false alarm.  This
possibility could also be checked by obtaining high-quality light
curves.  However, a failure to confirm an event predicted by the
HATNet photometric ephemeris may be inconclusive if the ephemeris is
not accurate enough, which often happens when considerable time has
elapsed between the HATNet observations and the follow-up work.
Absence of evidence may not imply evidence of absence.

\subsubsection{Evolved Primaries}

For seven of the candidates (HTR205-002, HTR205-006, HTR205-008,
HTR205-015, HTR205-016, HTR205-019, and HTR205-025) our analysis of
the CfA spectra implies that the primary star is evolved ($\log g =
3.25$ or weaker).  In all seven cases the available
proper motions are small, supporting the interpretation that the
targets are distant and therefore intrinsically luminous.  Of course,
a small proper motion is not proof that the distance to a star is
large, because the space motion of the star may be mostly directed
along the line of sight.  Conversely, a large proper motion does not
prove that the star is nearby, because metal-poor giants, although
quite rare, can have very large space motions.

Main Sequence stars come in a wide range of sizes, so it is not hard
to find a dwarf star with the right size to reproduce the observed
eclipse depth for each of the seven giants.  However, this is an
unlikely explanation, because the photometric periods are all very
short and would require that the companion orbit inside the giant.
Thus the possible explanations for the source of the dips in the light
curves of these giants are the same as for the two long-period
binaries discussed above (HTR205-019 and HTR205-021).  Indeed, one of
the seven giants is the primary in one of those binaries (HTR205-019),
with an orbital period of 94 days.

\subsubsection{Rapid Rotators}

For four of the candidates (HTR205-005, HTR205-007W, HTR205-009, and
HTR205-014) the line broadening observed in the CfA spectra implies
that the stars are rotating much too rapidly to allow radial
velocities to be determined with enough precision to derive a
spectroscopic orbit for a planetary companion.  As mentioned above,
the spectrum of HTR205-014 appears featureless in the Mg b region of
the spectrum, and no reliable radial velocity could be determined from
the CfA spectrum.  In the other three cases our analysis did yield
velocities, but with very large uncertainties.  The rotation that we
derived for HTR205-009 is 120 \kms, near the limit where the CfA
velocities become unreliable.  Therefore we did not bother to get a
second spectrum for this star.  For HTR205-005 and HTR205-007W we did
obtain additional spectra.  In both cases the resulting velocities
showed significant variation.  Presumably both stars are spectroscopic
binaries.

\subsubsection{Stars with No CfA Velocity Variations}

For seven of the unevolved stars, the CfA spectra yielded rotational
velocities slower than 10 \kms\ and $P(\chi^2)$ values for the velocity
residuals larger than 0.05, namely there was no evidence of velocity
variations.  Thus these seven systems survived the initial
reconnaissance spectroscopy and deserved further consideration as
viable planet candidates.  One of these seven is HTR205-001, which has a
very blue 2MASS color of $J-K_{\rm s} = 0.08$ mag.  Indeed the
effective temperature derived from the CfA spectra proved to be
uncomfortably hot, $T_{\rm eff} = 10,000$ K.  Nevertheless, the
rotation derived from the CfA spectra is quite modest, $v \sin i = 8$
\kms, and the three CfA velocities have an rms value of only 0.43
\kms.  If the source of the observed dip in the light curve is a
transiting planet, it would have to be enormously bloated.  Although
this candidate has not yet been absolutely ruled out as a transiting
planet, so far we have not followed up with additional observations.

The eastern star in the close visual pair HTR205-007 also survives as
a viable planet candidate.  In this case we have not pursued the
candidate with additional observations, partly because of the
difficulty in obtaining a high-quality light curve with KeplerCam for
a pair of stars that are separated by only $3\farcs2$.
Furthermore, the fact that the nearby companion appears to be a
spectroscopic binary with a large velocity amplitude and rapid
rotation suggests that it is likely to be the source of the dip in the
HATNet light curve.

Two of the surviving candidates, HTR205-026 and HTR205-030, were
subsequently withdrawn as photometric false alarms, based on
an independent reduction and analysis of their HATNet photometric time series
with the most recent tools and procedures developed by the HATNet
team.  It is interesting to note that the same reanalysis concluded
that six of the candidates classified spectroscopically as giants
(HTR205-002, HTR205-006, HTR205-008. HTR205-016, HTR205-019, and
HTR205-025) were also photometric false alarms in the original analysis.

Both members of the visual pair HTR205-022 and HTR205-023 with
separation $11\farcs2$ showed similar velocities and no significant
velocity variation based on the CfA spectra.  Subsequent follow-up
observations showed that HTR205-023 is the star hosting a transiting
planet \citep{Bakos:07b}, not HTR205-022.  The mean velocities and 
stellar parameters for the two stars are consistent with the interpretation
that they are a bound physical pair.

In this paper we report the confirmation of a planet transiting the
final of these seven candidates, HTR205-024, which we designate
\hatcur{b}.

\section{\hatcur{b}}

\subsection{HATNet Photometry}

\begin{figure}[!ht]
\plotone{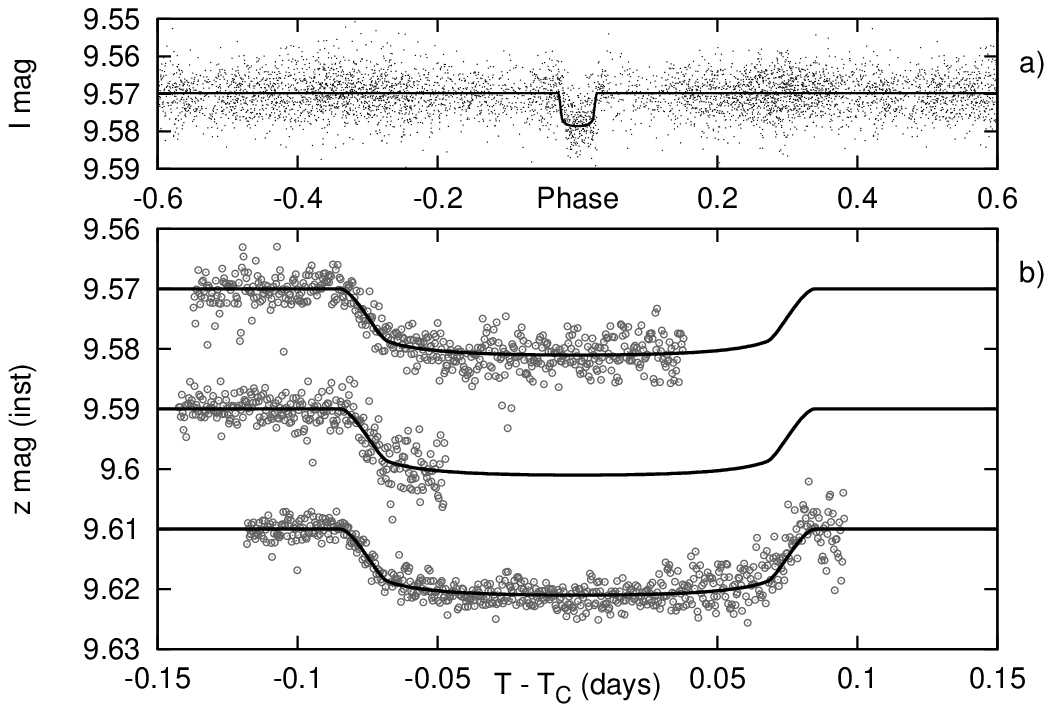}
\caption{
(a) The unbinned \band{I} \lc{} of \hatcur{} combined from four HATNet
telescopes, with 4460 points phased to the ephemeris derived in
\S~\ref{sec:lcrvanalysis}, with period $P = \hatcurLCPprec$\,days.  
The superimposed curve shows the best model fit using quadratic limb
darkening.
(b) The unbinned Sloan \band{z} light curves from two partial and one
complete transit events, acquired with KeplerCam on the \flwof{}
telescope on 2007 October 29, November 1 and December 2. The best-fit
transit model is superimposed.
\label{fig:lc}}
\end{figure}

In a previous publication \citep{Bakos:07a}, HTR205-023 was announced
as the first confirmed transiting planet from the HATNet survey and was
assigned the designation HAT-P-1b.  In this paper we confirm that a
second candidate in field G205, HTR205-024, is also a transiting
planet, which we designate HAT-P-8b.  This target is the
$I\approx\hatcurCCmag$ magnitude star \hatcurCCgsc{} (also known as
\hatcurCCtwomass{}; $\alpha = \hatcurCCra$, $\delta = \hatcurCCdec$;
J2000), with a transit depth of $\sim\hatcurLCdip$\,mmag, a period of
$P=\hatcurLCPshort$\,days and a relative duration (first to last
contact) of $q\approx0.049$, equivalent to a duration of $P \times q
\approx 3.6$~hours.  The HATNet \lc{} for \hatcur{} is plotted in
Figure \ref{fig:lc}.

\subsection{Highly Precise Velocities}

The initial spectroscopic reconnaissance of HTR205-024 with the CfA
Digital Speedometer on the 1.5m Tillinghast Reflector at FLWO on Mount
Hopkins yielded 8 spectra over a period of 357 days and a mean
velocity of $-22.53 \pm 0.79$ \kms\ rms and $P(\chi^2) = 0.18$.  Grids
of correlations against our library of calculated synthetic spectra
gave the best match for a template with $T_{\rm eff} = 6250$ K, $\log
g = 4.0$, and $v \sin i = 12.8$ \kms.  Encouraged by these positive
results, we scheduled HTR205-024 for high-quality spectroscopic
observations with HIRES \citep{Vogt:94} on the Keck~I telescope.
Between 2007 August 24 and September 1 we obtained 10 exposures with
an iodine cell, plus one iodine-free template, followed by two
additional iodine observations on 2007 October 23 and November 23. The
width of the spectrometer slit was $0\farcs86$, resulting in a
resolving power of $\lambda/\Delta\lambda \approx 55,\!000$, while the
wavelength coverage was $\sim3800-8000$\,\AA\@. The iodine gas
absorption cell was used to superimpose a dense forest of
$\mathrm{I}_2$ lines on the stellar spectrum in order to establish an
accurate wavelength fiducial \citep[see][]{Marcy:92}. Relative radial
velocities in the Solar System barycentric frame were derived as
described by \citet{Butler:96}, incorporating full modeling of the
spatial and temporal variations of the instrumental profile. The final
radial-velocity data and their errors are listed in
Table~\ref{tab:rvs}.  The velocity curve for our orbital solution (see
\S~\ref{sec:lcrvanalysis}) is plotted in Figure~\ref{fig:rvbis}a,
together with the observed velocities.

\subsection{Photometric Follow-up Observations}
To provide a high-quality light curve for \hatcur{}, KeplerCam on the
\mbox{1.2 m} telescope at FLWO was used to monitor two partial transit
events and one complete event, on 2007 October 29, November 1, and
December 2 UT, respectively; 526, 288, and 637 images in the Sloan
\band{z} were obtained, all with exposure times of 15 seconds and
cadence of 28 seconds.  After correction of the images using bias and
flat-field frames in the usual way, we derived a first order
astrometric transformation between the KeplerCam positions of
$\sim240$ stars and the corresponding 2MASS positions, as described in
\citet{Pal:06}, yielding rms residuals of $\sim0.4$ pixel.  Using
these positions we then performed aperture photometry using a series
of apertures with radii of 4.0, 6.0, and 8.0 pixels for the first
night, 4.5, 6.0, and 8.5 pixels for the second, and 10, 12, and 14
pixels for the third. Instrumental magnitude transformations were then
obtained using $\sim270$ stars on a frame taken near
culmination. First we derived a transformation that was
weighted by
the estimated Poisson and background noise of each star.
This led to an initial \lc\ and its associated rms for each star.
We then performed the magnitude transformation again, this time
weighting with the observed rms of the individual \lcs.  For each night
we chose the aperture that delivered the smallest rms photometric
residuals for the out-of-transit (OOT) observations of \hatcur{}; 8,
8.5, and 12 pixels, respectively. The resulting \lcs{} were then
de-correlated against trends based on a simultaneous fit of the light
curve parameters and the parameters of the possible trends (see
\S~\ref{sec:lcrvanalysis} for further details).  This yielded light
curves with an overall rms of $\sim2.3$\,mmag, for the first two
nights and $\sim1.9$\,mmag for the third night. These rms values are
slightly larger than the predicted value of 1.4\,mmag based on the
photon noise (1.1\,mmag) combined with the scintillation noise
\citep[0.8\,mmag at airmass 1.3; see][]{Dravins:98}. The excess noise suggests
that there may be modest residual trends that have not been
removed. The \lcs{} are shown in Figure~\ref{fig:lc}b.
The time series for the HATNet and KeplerCam photometry are published
in full in the electronic version of this paper.  The content and form
of these data are illustrated in Tables \ref{tab:hatnet} and
\ref{tab:keplercam}.

\ifthenelse{\boolean{emulateapj}}{
        \begin{deluxetable}{ccc}
}{
        \begin{deluxetable}{ccc}
}
\tablewidth{0pc}
\tablenum{9}
\tablecaption{HATNet Photometry for HTR205-024}
\label{tab:hatnet}
\tablehead{
\colhead{BJD} &
\colhead{Mag} &
\colhead{Error}}
\startdata
2453939.70260 & 9.57255 &  0.00461  \\
2453939.70642 & 9.56319 &  0.00372  \\
2453939.71020 & 9.56735 &  0.00358  \\
2453939.71398 & 9.57243 &  0.00354  \\
2453939.71777 & 9.56314 &  0.00364
\enddata
\tablecomments{
column (1): Barycentric Julian Day, \\
column (2): Best detrended HATNet magnitude, \\
column (3): Estimated error in the best magnitude. \\
Table 9 is presented in its entirety in the electronic edition
of the Astrophysical Journal. A portion is shown here for guidance
regarding its form and content.}
\ifthenelse{\boolean{emulateapj}}{
        \end{deluxetable}
}{
        \end{deluxetable}
}


\ifthenelse{\boolean{emulateapj}}{
        \begin{deluxetable}{cll}
}{
        \begin{deluxetable}{cll}
}
\tablewidth{0pc}
\tablenum{10}
\tablecaption{KeplerCam Photometry for HTR205-024}
\label{tab:keplercam}
\tablehead{
\colhead{BJD} &
\colhead{Mag} &
\colhead{Error}}
\startdata
2454403.69847 & $ 0.00311 $ &  0.00101 \\
2454403.69883 & $ 0.00133 $ &  0.00104 \\
2454403.69916 & $ 0.00580 $ &  0.00107 \\
2454403.69949 & $ 0.00011 $ &  0.00103 \\
2454403.69984 & $-0.01672 $ &  0.00111
\enddata
\tablecomments{
column (1): Barycentric Julian Day, \\
column (2): Best detrended KeplerCam $z$ magnitude, normalized to 0.0 out of transit, \\
column (3): Estimated error in the best magnitude.
Table 10 is presented in its entirety in the electronic edition
of the Astrophysical Journal. A portion is shown here for guidance
regarding its form and content.}
\ifthenelse{\boolean{emulateapj}}{
        \end{deluxetable}
}{
        \end{deluxetable}
}


\section{ANALYSIS}
\label{sec:analysis}

In this section we describe briefly our analysis yielding the orbital,
planetary, and stellar parameters of the \hatcur{} system.

\subsection{Light Curve and Radial-Velocity Analysis}
\label{sec:lcrvanalysis}

The analysis of the available photometric and radial-velocity data was
performed as follows.  First we tested whether the orbit could be
treated as circular, and fitted an eccentric orbit to the Keck/HIRES
RV measurements by adjusting the Lagrangian orbital elements
$k=e\cos\varpi$ and $h=e\sin\varpi$, along with the off-set of the
Keck/HIRES velocity system, (which is relative to the template
exposure), and the velocity semi-amplitude, $K$.  The orbital period
was fixed to the value found from the HATNet \lcs. We found that $k$
and $h$ are zero within nearly 1-$\sigma$, namely $k=0.05\pm0.04$,
$h=-0.01\pm0.03$, justifying our assumption of a circular orbit for
the subsequent analysis.

Following this, a joint fit was done using all of the available data,
including the HATNet and the follow-up light curves, and the Keck radial velocities, i.e.~we
adjusted the transit timings, light-curve parameters and the orbital
parameters simultaneously to achieve the best fit to the entire
data-set.  For the light curve modeling, we assumed quadratic stellar
limb darkening and we used the analytic formalism provided by
\citet{Mandel:02}. The appropriate limb darkening coefficients
$\gamma_1^{(z)}$, $\gamma_2^{(z)}$, $\gamma_1^{(I)}$ and
$\gamma_2^{(I)}$ were derived using the stellar atmospheric parameters
(see \S~\ref{sec:stellarparameters} for further details). The free
parameters were the \lc{} and orbital parameters, supplemented with
the parameters of the possible systematics.  Namely, the \lc{}
parameters were $T_{\mathrm{c},-485}$, the time of first transit
center in the HATNet campaign, $T_{\mathrm{c},11}$, the time of the
last observed transit center (on December 2), $m$, the out-of-transit
magnitude of the HATNet \lc{} in \band{I}, the fractional planetary
radius $p\equiv R_{\rm p}/R_\star$, the square of the impact parameter
$b^2$ and the quantity
$\zeta/R_\star=(2\pi/P)\cdot(a/R_\star)/\sqrt{1-b^2}$, while the
orbital parameters were the semi-amplitude of the radial velocity $K$
and the center-of-mass velocity $\gamma$ on the Keck system of relative velocities. The parameters of the
systematics were the terms linear in airmass and hour angle (which are
in practice equivalent with a linear and a quadratic term in time).
The $\zeta/R_\star$ parameter is related to the duration of the
transit as $(\zeta/R_\star)^{-1}=H$ where $2H$ is the time between the
instants when the center of the planet crosses the limb of the star
(i.e.~$2H$ is approximately the ``full-width at half maximum'' of the
transit, and is somewhat smaller than $T_{14}$, which is the time
between the first and last contact). For circular orbits, the impact
parameter is related to the orbital inclination $i$ as
$b\equiv(a/R_\star)\cos i$.  The parameters $\zeta/R_\star$ and $b^2$
were chosen instead of $a/R_\star$ and $b$ because their correlations
are negligible \citep[see][]{Pal:08b}. We note that in this joint fit
\emph{all} of the transits in the HATNet \lc{} have been adjusted
simultaneously, constrained by the assumption of a strictly periodic
signal. The shape of all these transits was characterized by $p$,
$b^2$ and $\zeta/R_\star$ (and the limb-darkening coefficients) while
the distinct transit center time instants were interpolated using
$T_{\mathrm{c},-485} =T_{\mathrm{c,first}}$ and
$T_{\mathrm{c},11}=T_{\mathrm{c,last}}$. For the initial values we
used the values provided by the BLS analysis, and the results provided
by the fit of a sinusoidal function for the folded and phased RV
data. We employed the method of refitting to synthetic data sets for
error estimation which gives the \emph{a posteriori} distribution of
the adjusted values. For each individual $\chi^2$ minimization, the
downhill simplex algorithm has been utilized \citep[a.k.a.~``AMOEBA'',
see][]{Press:92}.  The resulting parameter distribution was then used
directly as an input to the stellar evolution modeling, discussed
later in \S~\ref{sec:stellarparameters}.  We note that the joint fit
also yielded the period of the planetary orbit, namely
$P=(T_{\mathrm{c},11}-T_{\mathrm{c},-485})/496$, where 496 is the
number of cycles between the first and last observed transit events.

%
\ifthenelse{\boolean{emulateapj}}{
        \begin{deluxetable*}{lrrrrr}
}{
        \begin{deluxetable}{lrrrrr}
}
\tablewidth{0pc}
\tablenum{11}
\tablecaption{Relative Radial-Velocity Measurements
of \hatcur{}}
\label{tab:rvs}
\tablehead{
        \colhead{BJD}                           &
        \colhead{Phase}                         &
        \colhead{RV}                            &
        \colhead{\ensuremath{\sigma_{\rm RV}}}  &
        \colhead{BS}                            &
        \colhead{\ensuremath{\sigma_{\rm BS}}}\\ 
        \colhead{\hbox{~~~~(2,454,000$+$)~~~~}} &
        \colhead{$(-33+$}                       &
        \colhead{(\ms)}                         &
        \colhead{(\ms)}                         &
        \colhead{(\ms)}                         &
        \colhead{(\ms)}                          
}
\startdata
336.83041 \df &  0.2188 &$ -240.9 $& 4.3 &$  -3.8 $& 12.9  \\
336.83906 \df &  0.2216 &$    0.0 $& \nd &$ -11.1 $& 13.7  \\
337.79687 \df &  0.5329 &$  -47.4 $& 3.9 &$  -4.3 $& 12.8  \\
337.91459 \df &  0.5712 &$  -19.4 $& 3.8 &$  -1.2 $& 12.8  \\
338.80841 \df &  0.8618 &$  +27.0 $& 3.6 &$  -0.9 $& 12.8  \\
338.91425 \df &  0.8962 &$    0.0 $& 3.6 &$  +8.5 $& 11.9  \\
339.79280 \df &  1.1818 &$ -243.7 $& 3.4 &$  -1.5 $& 12.8  \\
343.86602 \df &  2.5058 &$ -100.9 $& 4.4 &$  +8.8 $& 11.9  \\
344.99747 \df &  2.8736 &$  +21.4 $& 4.6 &$  +0.3 $& 12.4  \\
396.80973 \df & 19.7159 &$  +49.3 $& 5.0 &$  +4.8 $& 12.5  \\
427.80583 \df & 29.7916 &$  +61.6 $& 5.5 &$  +0.2 $& 12.8    
\enddata
\ifthenelse{\boolean{emulateapj}}{
        \end{deluxetable*}
}{
        \end{deluxetable}
}

\begin{figure} 
\plotone{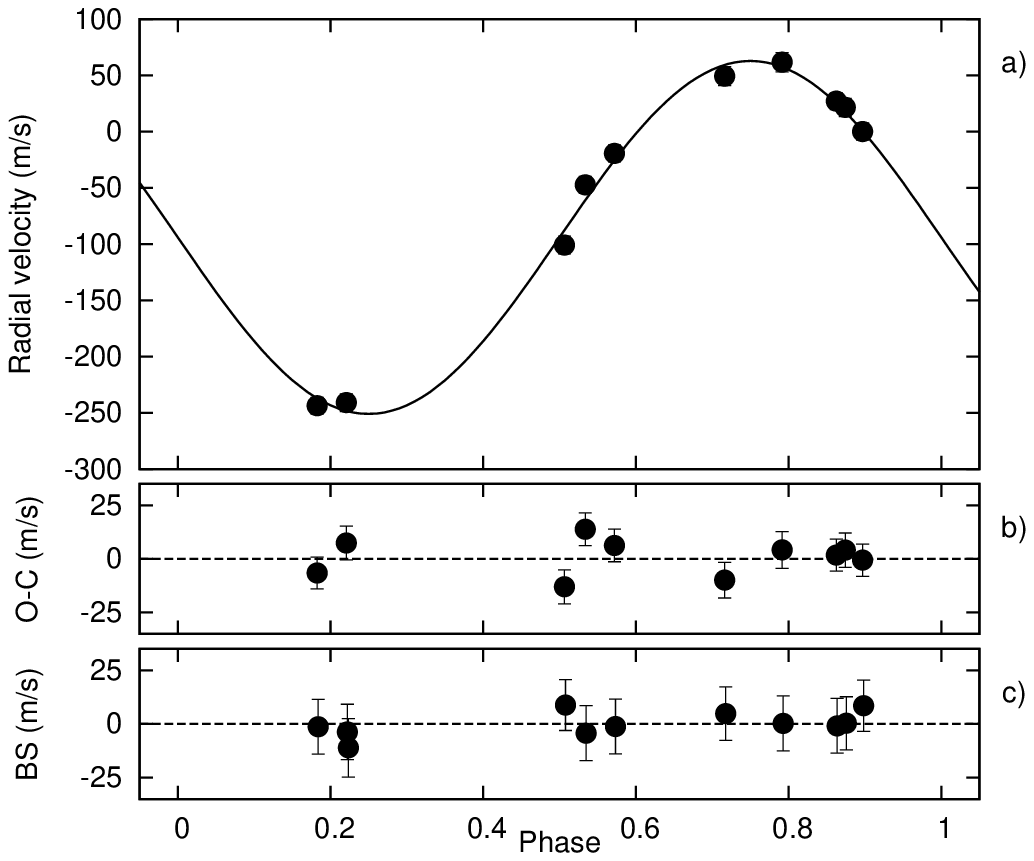}
\caption{
	a) The velocity curve for \hatcur{} and the observed radial
	velocities obtained with Keck, using our orbital
	solution with the best period (see
	\S~\ref{sec:lcrvanalysis}). The zero point of the radial
	velocities is relative to the Keck template spectrum.
	b) The velocity residuals from the orbital solution (also see
	\S~\ref{sec:lcrvanalysis}). The rms variation of the residuals is
	about 3.8\,\ms.
	c) The bisector spans (BS) for the 10 Keck spectra plus the
	single template spectrum, computed as described in the text.
	The mean value has been subtracted. Due to the relatively
	small errors compared to the radial-velocity amplitude, the
	vertical scale in the b) and c) panels differs from the scale
	used in panel a).
\label{fig:rvbis}}
\end{figure}

Using the distribution of these parameters, it is straightforward to
obtain the values and the errors of the additional derived parameters,
e.g. $a/R_\star=(\zeta/R_\star)\sqrt{1-b^2}(P/2\pi)$. 

\begin{deluxetable}{lcl}
\tablewidth{0pc}
\tablenum{12}
\tablecaption{Stellar Parameters for \hatcur{}}
\label{tab:stellar}
\tablehead{\colhead{Parameter}	& \colhead{Value} 		& \colhead{Source}}
\startdata
$\teffstar$ (K)\dotfill			&  \hatcurSMEteff		& SME\tablenotemark{a} 	\\
$[\mathrm{Fe/H}]$\dotfill		&  \hatcurSMEzfeh		& SME					\\
$v \sin i$ (\kms)\dotfill		&  \hatcurSMEvsin		& SME					\\
$M_\star$ ($M_{\sun}$)\dotfill  &  \hatcurYYm			& Y$^2$+LC+SME\tablenotemark{b}	\\
$R_\star$ ($R_{\sun}$)\dotfill  &  \hatcurYYr			& Y$^2$+LC+SME			\\
$\log g_\star$ (cgs)\dotfill    &  \hatcurYYlogg		& Y$^2$+LC+SME			\\
$L_\star$ ($L_{\sun}$)\dotfill  &  \hatcurYYlum			& Y$^2$+LC+SME			\\
$M_V$ (mag)\dotfill				&  \hatcurYYmv   		& Y$^2$+LC+SME			\\
Age (Gyr)\dotfill				&  \hatcurYYage			& Y$^2$+LC+SME			\\
Distance (pc)\dotfill			&  \hatcurXdist			& Y$^2$+LC+SME
\enddata
\tablenotetext{a}{SME = `Spectroscopy Made Easy' package for analysis
of high-resolution spectra \cite{Valenti:96}. See text.}
\tablenotetext{b}{Y$^2$+LC+SME = Yale-Yonsei isochrones \citep{Yi:01},
\lc{} parameters, and SME results.}
\end{deluxetable}

\subsection{Stellar Parameters}
\label{sec:stellarparameters}

We used the iodine-free template spectrum from Keck for an initial
determination of the atmospheric parameters. Spectral synthesis
modeling was carried out using the SME software \citep{Valenti:96},
with wavelength ranges and atomic line data as described by
\citet{Valenti:05}. We obtained the following initial values: effective
temperature $6050\pm70$\,K, surface gravity $\log g_\star =
3.97\pm0.10$ (cgs), iron abundance $\mathrm{[Fe/H]}=-0.08\pm0.05$, and
projected rotational velocity $v \sin i=11.6\pm0.5$\,\kms.  The
uncertainties quoted here and in the remainder of this discussion are
approximately twice the statistical uncertainties for the values given
by the SME analysis. This reflects our attempt, based on prior
experience, to account for possible systematic errors (e.g. \cite{Noyes:08}; see
also \cite{Valenti:05}).  Note that the previously discussed limb
darkening coefficients, $\gamma_1^{(z)}$, $\gamma_2^{(z)}$,
$\gamma_1^{(I)}$ and $\gamma_2^{(I)}$ have been taken from the tables
of \cite{Claret:04} by interpolation to the above-mentioned SME values
for $\teffstar$, $\log g_\star$, and $\mathrm{[Fe/H]}$.

The results of the joint fit, together with the initial results from
spectroscopy enable us to refine the parameters of the star.  As
described by \citet{Sozzetti:07} and \citet{Torres:08}, $a/R_\star$ is
a better luminosity indicator than the spectroscopic value of $\log
g_\star$, because the variation of stellar surface gravity has a subtle
effect on the line profiles. Therefore, we used the values of
$\teffstar$ and $\mathrm{[Fe/H]}$ from the initial SME analysis,
together with the distribution of $a/R_\star$ to estimate the stellar
properties from comparison with the Yonsei-Yale (Y$^2$) stellar
evolution models by \cite{Yi:01}. As was discussed in
\S~\ref{sec:lcrvanalysis}, a Monte-Carlo set for $a/R_\star$ values has
been generated by the joint fit. We used this distribution for stellar
evolution modeling as described in \citet{Pal:08a}. The set of the
\emph{a posteriori} distribution of the stellar parameters was
therefore obtained, including the mass, radius, age, luminosity and
color (in multiple bands). Since the mass and radius (and their
respective distributions) of the star are known, it is straightforward
to obtain the surface gravity and its uncertainty together. The derived
surface gravity was $\log g_\star=4.19^{+0.02}_{-0.04}$, which is
somewhat larger than the previous value provided by the SME analysis.
Therefore, we repeated the atmospheric modeling by fixing the surface
gravity and letting only the other parameters vary. The next iteration
of the SME analysis resulted in the following values: effective
temperature $\teffstar=\hatcurSMEteff$\,K, metallicity
$\mathrm{[Fe/H]}=\hatcurSMEzfeh$, and projected rotational velocity $v
\sin i=\hatcurSMEvsin$\,\kms. Based on these new spectroscopic values,
we updated the limb darkening coefficients and repeated the light curve
and radial-velocity simultaneous fit (except for the systematics
parameters which were fixed here) plus the stellar evolution modeling
in the same way as described in \S~\ref{sec:lcrvanalysis} and earlier
in this subsection.  The resulting stellar surface gravity was $\log
g_\star=\hatcurYYlogg$, which is well within 1-$\sigma$ of the value
obtained in the previous iteration. Therefore we accept the values from
the second joint fit and stellar evolution modeling as the final light
curve and stellar parameters (see Table~\ref{tab:stellar}).

The Yonsei-Yale isochrones also contain the absolute magnitudes and
colors for different photometric bands from $U$ up to $M$, providing an
easy comparison of the estimated and the observed magnitudes and
colors.  Using these data, we determined the $V-I$ model color,
$(V-I)_{\rm YY}=\hatcurYYvi$. We have compared the $(V-I)_{\rm YY}$
color to published observational data, and found that our model color 
agrees well with
the observed TASS color of $(V-I)_{\rm TASS}=\hatcurCCtassvi$
\citep[see][]{Droege:06}.  Hence, the star is not affected by 
interstellar reddening within the errors, since $E(V-I)\equiv(V-I)_{\rm
TASS}-(V-I)_{\rm YY}=0.04\pm0.10$ (the galactic latitude of \hatcur{}
is $b=-21\arcdeg22\arcmin$). For estimating the distance of \hatcur{},
we used the absolute magnitude $M_V=\hatcurYYmv$ (resulting from the
isochrone analysis, see also Table~\ref{tab:stellar}) and the $V_{\rm
TASS}=\hatcurCCtassmv\pm0.07$ observed magnitude. These two yield a
distance modulus of $V_{\rm TASS}-M_V=6.82\pm0.14$, i.e.~a distance of
$d=\hatcurXdist$\,pc.

\subsection{Planetary and Orbital Parameters}
\label{sec:planetparams}

As described in \citet{Pal:08a}, the planetary parameters and
their uncertainties can be derived by the direct combination of the
\emph{a posteriori} distributions of the light curve, radial-velocity
and stellar parameters.  We found that the mass of the planet is
$M_p=\hatcurPPm$\,\mjup, the radius is $R_p=\hatcurPPr$\,\rjup\
and its density is $\rho_p=\hatcurPPrho$\,\gcmc. The final planetary
parameters are also summarized at the bottom of
Table~\ref{tab:parameters}.

\subsection{Excluding Blend Scenarios}

Following \cite{Torres:07}, we explored the possibility that the
measured radial velocities are not real, but instead are caused by
distortions in the spectral line profiles due to contamination from an
unresolved eclipsing binary.  In that case the ``bisector span'' of
the average spectral line should vary periodically with amplitude and
phase similar to the measured velocities themselves
\citep{Queloz:01,Mandushev:05}. We cross-correlated each Keck spectrum
against a synthetic template matching the properties of the star
(i.e.~based on the SME results, see \S~\ref{sec:stellarparameters}),
and averaged the correlation functions over all orders blueward of the
region affected by the iodine lines. From this representation of the
average spectral line profile we computed the mean bisectors, and as a
measure of the line asymmetry we computed the ``bisector spans'' as
the velocity difference between points selected near the top and
bottom of the mean bisectors \citep{Torres:05}. If the velocities were
the result of a blend with an eclipsing binary, we would expect the
line bisectors to vary in phase with the photometric period with an
amplitude similar to that of the velocities. Instead, we detect no
variation in excess of the measurement uncertainties (see
Fig.~\ref{fig:rvbis}c). Therefore, we conclude that the velocity
variations are real and that the star is orbited by a Jovian
planet. We note here that the mean bisector span ratio relative to the
radial-velocity amplitude is the smallest ($\sim 0.026$) among all the
HATNet planets, indicating an exceptionally high confidence that the
RV signal is not due to a blend with an eclipsing binary companion.

\begin{deluxetable}{lc}
\tablewidth{0pc}
\tablenum{13}
\tablecaption{Orbital and planetary parameters}
\label{tab:parameters}
\tablehead{\colhead{~~~~~~~~~~~~~~~Parameter~~~~~~~~~~~~~~~} & \colhead{Value}}
\startdata
\sidehead{\Lc{} parameters}
~~~$P$ (days)                         \dotfill            & $\hatcurLCP$         \\
~~~$E$ (${\rm BJD}-2,\!400,\!000$)    \dotfill            & $\hatcurLCMT$        \\
~~~$T_{14}$ (days)\tablenotemark{a}   \dotfill            & $\hatcurLCdur$       \\
~~~$T_{12} = T_{34}$ (days)\tablenotemark{a} \dotfill     & $\hatcurLCingdur$    \\
~~~$a/R_\star$                        \dotfill            & $\hatcurPPar$        \\
~~~$R_p/R_\star$                      \dotfill            & $\hatcurLCrprstar$   \\
~~~$b \equiv a \cos i/R_\star$        \dotfill            & $\hatcurLCimp$       \\
~~~$i$ (deg)                          \dotfill            & $\hatcurPPi$ \phn    \\
\sidehead{Spectroscopic parameters}
~~~$K$ (\ms)                          \dotfill            & $\hatcurRVK$         \\
~~~$\gamma$ (\kms)                    \dotfill            & $22.53 \pm 0.28$     \\
~~~$e$                                \dotfill            & $0$ (adopted)        \\
\sidehead{Planetary parameters}
~~~$M_p$ ($\mjup$)                    \dotfill            & $\hatcurPPm$         \\
~~~$R_p$ ($\rjup$)                    \dotfill            & $\hatcurPPr$         \\
~~~$C(M_p,R_p)$                       \dotfill            & $\hatcurPPmrcorr$    \\
~~~$\rho_p$ (\gcmc)                   \dotfill            & $\hatcurPPrho$       \\
~~~$a$ (AU)                           \dotfill            & $\hatcurPParel$      \\
~~~$\log g_p$ (cgs)                   \dotfill            & $\hatcurPPlogg$      \\
~~~$T_{\rm eq}$ (K)                   \dotfill            & $\hatcurPPteff$      \\
~~~$\Theta$                           \dotfill            & $\hatcurPPtheta$
\enddata
\tablenotetext{a}{\ensuremath{T_{14}}: total transit duration, time
between first to last contact; \ensuremath{T_{12}=T_{34}}:
ingress/egress duration, namely the times between first and second, or
third and fourth contact.}
\end{deluxetable}


\section{DISCUSSION}  
\label{sec:discussion}

In this paper we describe the procedures and observations that we use
to follow up transiting-planet candidates identified by wide-angle
ground-based photometric surveys, and we report the status of our
efforts to follow up 32 candidates identified in HATNet field G205.

Initial spectroscopic observations obtained with the CfA Digital
Speedometers disclosed that twelve of the candidates showed variable
radial velocities with amplitudes of at least a few \kms, indicating
that in each case the companion responsible for the velocity
variations must be too massive to be a planet.  We accumulated enough
observations for ten of these spectroscopic binaries to allow the
derivation of orbital solutions.  Eight of these orbits have
parameters that are consistent with the photometric ephemerides, thus
proving that the stellar companion is the source of the transit-like
light curve.  For all but two of these eight eclipsing binaries the
original HATNet photometric period turned out to be half of the true
orbital period, presumably because the primary and secondary eclipses
were not distinguishable in the HATNet photometry, implying that the two
stars must be similar.  Given the shallowness of the observed dips,
the eclipses must be grazing, a conclusion that is supported by the fact
that we were able to detect the spectrum of the secondary and derive
double-lined orbits for six of the seven.  For the two eclipsing
binaries where the orbital period matches the original HATNet
photometric period, the companions are M dwarfs, small enough to
produce a transit-like light curve with full eclipses of the F-star
primary but no detectable secondary eclipses \citep{Beatty:07}.

\begin{figure}[!ht]
\plotone{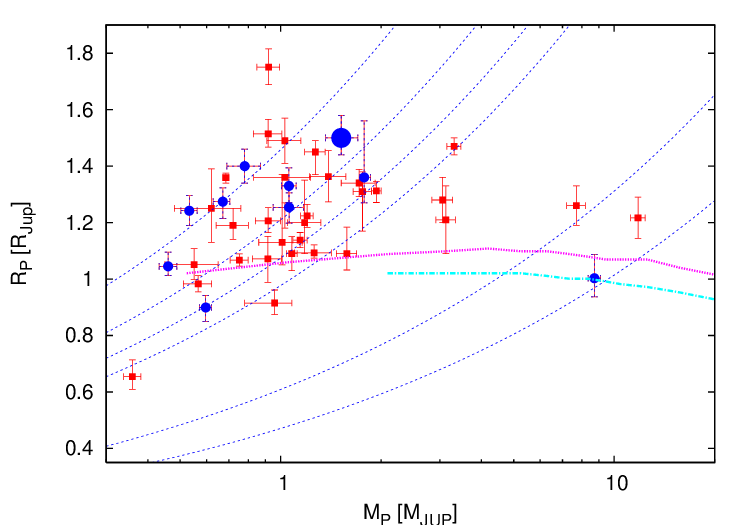}
\caption{ Mass--radius diagram for published transiting extrasolar
planets.  The data are taken from \citet{Torres:08}, and
www.exoplanet.eu.  Filled circles mark HATNet discoveries, and the
large filled circle marks \hatcur{b}. Overlaid are \citet{Baraffe:03}
(zero insolation planetary) isochrones for ages of 0.5~Gyr (upper,
dotted line) and 5~Gyr (lower dashed-dotted line),
respectively. \hatcur{b} lies well above the main stream for Hot
Jupiters.
\label{fig:mr}}
\end{figure}

Two of the single-lined spectroscopic binaries have orbital periods
considerably longer than the HATNet photometric periods.  Thus the
unseen companion responsible for the orbital motion can not be the
source of the transit-like light curve.  Possible explanations include
hierarchical triple systems with an eclipsing companion to either one
of the stars in the observed outer orbit, or a quadruple system with
an eclipsing binary in a much wider orbit than the one observed.  The
eclipsing binary could even be an accidental alignment, unresolved in
our images.  There is also the possibility that the original HATNet
detection was a photometric false alarm.

We also used the spectra from the CfA Digital Speedometers to estimate
the effective temperature and surface gravity of the candidate stars.
In seven cases our spectroscopic gravities implied that the stars are
giants, too large to allow detection of a transiting planet with
HATNet, let alone that with such short periods, such putative planets
would orbit inside the giant star. 
The possible explanations for these systems are similar to
those in the previous paragraph. Either the giant is diluting the
light of an eclipsing binary, or the photometric detection is a false
alarm.  Indeed, one of the giants is also the primary in one of the
two spectroscopic binaries discussed in the previous paragraph.

Four of the candidates are rotating too rapidly to allow very precise
radial-velocity measurements, and two of these also showed large
velocity variations based on the CfA spectra.  Thus these four
candidates were not pursued further.

Four of the candidates have not yet been followed up with initial
reconnaissance spectroscopy because they are fainter than $V=13.0$ mag
and difficult targets for the old CfA Digital Speedometers.  We plan
to follow them up in the near future with the Tillinghast Reflector
Echelle Spectrograph, a modern fiber-fed echelle spectrometer that has
recently come into operation at FLWO.

Spectra obtained with the CfA Digital Speedometers failed to show
velocity variations at the level of about 0.5 to 1.0 \kms\ for seven of the
candidates.  Two of these were subsequently withdrawn as photometric
false alarms.  One is quite hot and has not yet been followed up with
very precise radial velocities. One is a member of a close visual binary where the
companion is a spectroscopic binary that is likely to be the source of
the dip in the HATNet light curve.  Two are the components of another
visual binary, one of which was previously confirmed as the host of the
transiting planet HAT-P-1b.  In this paper we report that the seventh of
these candidates is also a planet, \hatcur{b}.

Our newly confirmed planet has mass $M_{\rm p} = \hatcurPPm\,\mjup$\ and
radius $R_{\rm p} = \hatcurPPr\,\rjup$, which places it among the most
inflated of the transiting Hot Jupiters currently known (see
Figure~\ref{fig:mr}).  The 3.3\,\lsun\ luminosity of the host star and
the $a=\hatcurPParel$\,AU semi-major axis correspond to an equivalent
semi-major axis of $a=0.026$\,AU if this object orbited our Sun and
received the same flux. (Note, however, that the spectrum of the solar
flux would be different).  The theoretical radius from
\citet{Fortney:08} would be 1.32\rjup, 1.25\rjup\ and 1.17\rjup,
respectively, for ages of 300\,Myr, 1\,Gyr and 4.5\,Gyr, assuming
coreless models, and 0.02\,AU orbital distance. \hatcur{b} with its
\hatcurPPr\,\rjup\ radius is thus inflated at the 2-3-$\sigma$
level. This is also consistent with \citet{liu:08}, where the
1.5\,\mjup\ mass and 0.026\,AU equivalent solar distance yields such a
large equilibrium radius only if the ratio between core heating power
and insolation power is as high as $2\times10^{-3}$. The equilibrium
time-scale for the radius evolution would be very short, approx
20\,Myr. For comparison, WASP-5b \citep{Anderson:08} has a similar mass
(1.58\,\mjup), but its radius is only 1.09\,\rjup.

\begin{figure}[!ht]
\plotone{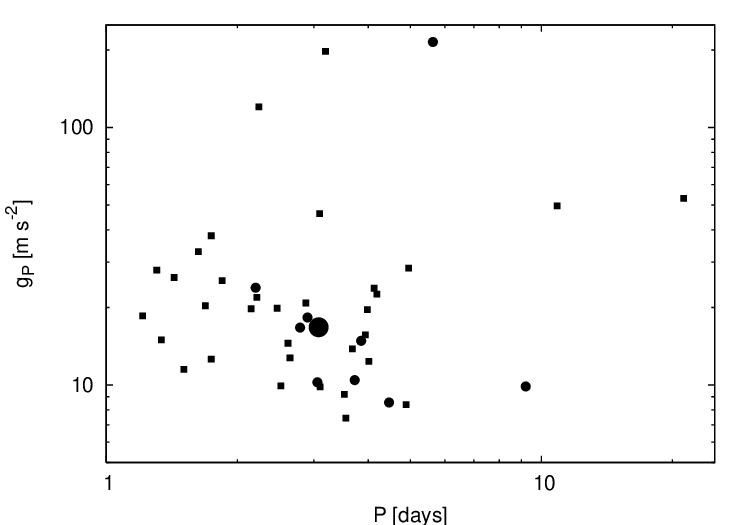}
\caption{ The period vs.~surface gravity diagram for published
transiting extrasolar planets. The data are taken from
\citet{Torres:08}, and www.exoplanet.eu. Filled circles mark HATNet
discoveries, and the large filled circle marks \hatcur{b}, which
lies on the main stream for Hot Jupiters, if the
supermassive planets, such as HAT-P-2b and XO-3b, are ignored.
\label{fig:pg}}
\end{figure}

\hatcur{b} is also offset to the high-mass side of the period--mass
relation. Planets with orbital periods similar to \hatcur{b} tend to
have smaller masses, around 0.8\,\mjup\ as compared to the
\hatcurPPmshort\,\mjup\ mass of \hatcur{b}.  However, with the
discovery of more massive planets, such as HAT-P-2b \citep{Bakos:07a},
XO-3b \citep{Johns-Krull:08}, and Corot-Exo-3b \citep{Deleuil:08} the
period--mass relation has become less clear-cut.  Although the mass
and radius of \hatcur{b} are larger than usual, together they imply
an unremarkable surface gravity, $\log g = \hatcurPPlogg$ (cgs), and
\hatcur{b} falls tightly on the period--surface gravity relations
reported by \citet{Southworth:07} and by \citet{Torres:08}.

The Safronov number of \hatcurb\ is $\Theta = \hatcurPPtheta$,
indicating that it belongs to the so-called Class-I planets as defined
by \citet{Hansen:07}. Curiously, \hatcur{b} lies at the very hot end
of the distribution in the equilibrium temperature vs.~Safronov number
diagram. All the hotter transiting planets are in Class
II\@. Furthermore, while ``inflated'' planets mostly belong to Class II,
\hatcur{b} has an inflated radius, yet belongs to Class I. With the
recent discovery of several new transiting exoplanets, the Safronov
dichotomy is becoming less pronounced, e.g.~WASP-11b/HAT-P-10b
\citep{West:08, Bakos:08} falls in between the two groups, and the
high- and low-mass exoplanets also seem to fall outside these groups.

The incident flux on \hatcur{b} is $1.91 \times 10^{9}$ erg s$^{-1}$
cm$^{-2}$, placing it in the pM class, as defined by
\citet{Fortney:08}. This implies that \hatcur{b} has significant
opacity due to the absorption by molecular TiO and ViO in its
atmosphere, leading to a temperature inversion and a hot
stratosphere. This could be tested with future {\em Spitzer Space
Telescope} \/observations. Based on its mass and incident flux,
\hatcur{b} is similar to TrES-3 \citep{O'Donovan:07}.

The Transiting Exoplanet Survey Satellite (TESS) is under study for a
NASA SMEX mission to survey the entire sky for nearby bright
transiting planets.  The TESS cameras are similar to the HATNet
cameras, with a similar pixel size and similar image quality.  Thus
TESS will share some of the same challenges that HATNet has faced in
following up transiting-planet candidates.  TESS will observe targets
selected from an input catalog, with the goal of avoiding evolved
stars.  This should eliminate one source of false positives that has
plagued the magnitude-limited ground-based photometric surveys.  In
the case of HATNet field G205, 8 out of 28 candidates proved to be
evolved stars.  In addition, the most interesting targets for TESS
will be the nearest and brightest stars, for example roughly 50,000
Main Sequence stars from the Hipparcos Catalogue.  Because these stars
have been much better studied over the years than the fainter stars
typical of the HATNet candidates, much more will be known about them
ahead of time.  TESS will achieve roughly ten times better limiting
photometric precision than HATNet, and the frequency of grazing
eclipsing binaries masquerading as transiting planets will be much
smaller for light curves with shallower dips, compared to the 7
grazing eclipsing binaries out of 28 candidates found in HATNet field
G205.  On the other hand, some of the shallower dips will be due to
fainter eclipsing binaries that contaminate the light of the primary
targets.  These will pose a difficult challenge in the follow-up of
TESS candidates.


\acknowledgements 

HATNet operations have been funded by NASA grants NNG04GN74G,
NNX08AF23G and SAO IR\&D grants Work by G.\'A.B.~was supported by NASA
through Hubble Fellowship Grant HST-HF-01170.01-A and by the
Postdoctoral Fellowship of the NSF Astronomy and Astrophysics Program.
We acknowledge partial support also from the Kepler Mission under NASA
Cooperative Agreement NCC2-1390 (D.W.L., PI). G.T.~acknowledges partial
support from NASA under grant NNG04LG89G, G.K.~thanks the Hungarian
Scientific Research Foundation (OTKA) for support through grant
K-60750. We thank for the help of F.~Pont, F.~Bouchy and A.~Shporer in
the initial spectroscopy of HTR205-024. This research has made use of
Keck telescope time granted through NOAO (program A285Hr).





\begin{thebibliography}{}

\bibitem[Anderson et al.(2008)]{Anderson:08}
 Anderson, D. R., Gillon, M., Hellier, C., Maxted, P. F. L., Pepe, F.,
 Queloz, D., Wilson, D. M., Collier Cameron, A. 2008, \mnras, 387, L4

\bibitem[Bakos et al.(2002)]{Bakos:02}
 Bakos, G.~\'A., L\'az\'ar, J., Papp, I., S\'ari, P.,
 \& Green, E.~M.~2002, \pasp, 114, 974

\bibitem[Bakos et al.(2004)]{Bakos:04}
 Bakos, G.~\'A., Noyes, R.~W., Kov\'acs, G., Stanek, K.~Z.,
 Sasselov, D.~D., \& Domsa, I.~2004, \pasp, 116, 266

\bibitem[Bakos et al.(2007)]{Bakos:07a} 
 Bakos, G.~\'A., Kov\'acs, G., Torres, G., Fischer, D. A., Latham,
 D. W., Noyes, R. W., Sasselov, D. D., Mazeh, T., et al.~2007a,
 \apj, 670, 826

\bibitem[Bakos et al.(2007)]{Bakos:07b}
 Bakos, G.~\'A., Noyes, R. W., Kov\'acs, G., Latham, D. W., Sasselov,
 D. D., Torres, G., Fischer, D. A., Stefanik, R. P., et al.~2007b,
 \apj, 656, 552

\bibitem[Bakos et al(2008)]{Bakos:08} 
 Bakos, G. \'A.; P\'al, A., Torres, G., Sip\H{o}cz, B., Latham, D. W.,
 Noyes, R. W., Kov\'acs, Geza, Hartman, J., Esquerdo, G. A., et al.
 2008, \apj\ submitted, arXiv:0809.4295

\bibitem[Baraffe et al.(2003)]{Baraffe:03}
 Baraffe, I., Chabrier, G., Barman, T.~S., Allard, F., \& Hauschildt,
 P.~H.~2003, \aap, 402, 701

\bibitem[Beatty et al.(2007)]{Beatty:07}
 Beatty, T.~G., Fe\'andez, J. M., Latham, D. W., Bakos, G.~\'A.,
 Kov\'acs, G., Noyes, R. W., Stefanik, R. P., Torres, G., et al.~2007,
 \apj, 663, 573

\bibitem[Brown et al.(2007)]{Brown:07}
 Brown, T.~M., Rosing, W. E., Baliber, N., Hidas, M., \& Street,
 R.~2007, \baas, 38, 173

\bibitem[Butler et al.(1996)]{Butler:96} 
 Butler, R.~P., Marcy, G. W, Williams, E., McCarthy, C., Dosanjh, P.,
 \& Vogt, S. S. 1996, \pasp, 108, 500

\bibitem[Butler et al.(2006)]{Butler:06} 
 Butler, R.~P., Wright, J. T., Marcy, G. W., Fischer, D. A., Vogt,
 S. S., Tinney, C. G., Jones, H. R. A., Carter, B. D., et al. 2006,
 \apj, 646, 505

\bibitem[Claret(2004)]{Claret:04}
 Claret, A.~2004, \aap, 428, 1001

\bibitem[Deleuil et al.(2008)]{Deleuil:08}
 Deleuil, M., Deeg, H. J., Alonso, R., Bouchy, F., \& Rouan, D.
 2008, \aap\ accepted, arXiv:0810.0919

\bibitem[Demarque et al.(2004)]{Demarque:04}
 Demarque, P., Wo, J.-H., Kim, Y. C., \& Yi, S. K. 2004, \apj, 155, 667

\bibitem[Dravins et al.(1998)]{Dravins:98}
 Dravins, D., Lindegren, L., Mezey, E., \& Young, A. T. 1998,
 \pasp, 110, 610

\bibitem[Droege et al.(2006)]{Droege:06}
 Droege, T.~F., Richmond, M.~W., Sallman, M. P., \& Creager,
 R. P. 2006, \pasp, 118, 1666

\bibitem[Fortney et al.(2008)]{Fortney:08}
 Fortney, J. J., Lodders, K., Marley, M. S., \& Freedman, R. S. 2008,
 \apj, 678, 1419

\bibitem[Hansen \& Barman(2007)]{Hansen:07}
 Hansen, B. M. S., \& Barman, T. 2007, \apj, 671, 861

\bibitem[Johns-Krull et al.(2008)]{Johns-Krull:08}
 Johns-Krull, C. M., McCullough, P. R., Burke, C. J., Valenti, J. A.,
 Janes, K. A., Heasley, J. N., Prato, L., Bissinger, R., et al.~2008,
 \apj, 677, 657

\bibitem[Kov\'acs et al.(2002)]{Kovacs:02}
 Kov\'acs, G., Zucker, S., \& Mazeh, T.~2002, \aap, 391, 369

\bibitem[Kov\'acs et al.(2005)]{Kovacs:05}
 Kov\'acs, G., Bakos, G.~\'A., \& Noyes, R.~W.~2005, \mnras, 356, 557

\bibitem[Latham(1992)]{Latham:92}
 Latham, D.~W. 1992, in IAU Coll.~135, Complementary Approaches to
 Double and Multiple Star Research, ASP Conf.~Ser.~32, 
 eds.~H.~A.~McAlister \& W.~I.~Hartkopf (San Francisco: ASP), 110

\bibitem[Latham et al.(2002)]{Latham:02}
 Latham, D. W., Stefanik, R. P., Torres, G., Davis, R. J., Mazeh, T.,
 Carney, B. W., Laird, J. B., \& Morse, J. A. 2002, \aj, 124, 1144

\bibitem[Latham(2003)]{Latham:03}
 Latham, D. W. 2003.  In Scientific Frontiers in Research on
 Extrasolar Planets, ASP Conf. Ser. 294, eds. D. Deming \& S. Seager
 (San Francisco: ASP), 409

\bibitem[Liu et al.(2008)]{liu:08}
 Liu, X., Burrows, A., \& Ibgui, L.~2008, \apj\ accepted, arXiv:0805.1733

\bibitem[Mandel \& Agol(2002)]{Mandel:02}
 Mandel, K., \& Agol, E.~2002, \apjl, 580, L171

\bibitem[Mandushev et al.(2005)]{Mandushev:05} 
 Mandushev, G., Torres, G., Latham, D. W., Charbonneau, D., Alonso,
 R., White, R. J., Stefanik, R. P., Dunham, E. W., ~et al.~2005, \apj,
 621, 1061

\bibitem[Marcy \& Butler(1992)]{Marcy:92}
 Marcy, G.~W., \& Butler, R.~P. 1992, \pasp, 104, 270

\bibitem[McCullough et al.(2006)]{McCullough:06}
 McCullough, P.~R., Stys, J. E., Valenti, J. A., Johns-Krull, C. M.,
 Janes, K. A., Heasley, J. N., Bye, B. A., Dodd, C., et al.~2006, \apj,
 648, 1228

\bibitem[Noyes et al.(2008)]{Noyes:08}
 Noyes, R.~W., Bakos, G.~\'A., Torres, G., P\'al, A., Kov\'acs, G.,
 Latham, D. W., Fern\'andez, J. M., Fischer, D. A., et al.~2008, \apjl,
 673, L79

\bibitem[O'Donovan et al.(2007)]{O'Donovan:07}
O'Donovan, F. T., Charbonneau, D., Bakos, G.~\'A., Mandushev, G.,
Dunham, E. W., Brown, T. M., Latham, D. W., Torres, G., et al. 2007.
\apjl, 663, L37

\bibitem[P\'al \& Bakos(2006)]{Pal:06}
 P\'al, A., \& Bakos, G.~\'A. 2006, \pasp, 118, 1474

\bibitem[P\'al et al.(2008)]{Pal:08a}
 P\'al, A., Bakos, G.~\'A., Torres, G., Noyes, R.~W., Latham, D. W.,
 Kov\'acs, G., Marcy, G. W., Fischer, D. A., et al.~2008, \apj, 680,
 1450

\bibitem[P\'al(2008)]{Pal:08b}
 P\'al, A. 2008, \mnras, 390, 281

\bibitem[Press et al.(1992)]{Press:92}
 Press, W. H., Teukolsky, S. A., Vetterling, W. T., \& Flannery, B. P., 1992,
 Numerical  Recipes in C: the art of scientific computing, Second Edition, Cambridge University Press

\bibitem[Queloz et al.(2001)]{Queloz:01}
 Queloz, D., Henry, G. W., Sivan, J. P., Baliunas, S. L., Beuzit,
 J. L., Donahue, R. A., Mayor, M., Naef, D., ~et al.~2001, \aap, 379,
 279

\bibitem[Skrutskie et al.(2006)]{Skrutskie:06}
 Skrutskie, M. J., Cutri, R. M., Stiening, R., Weinberg, M. D.,
 Schneider, S., Carpenter, J. M., Beichman, C., Capps, R., et
 al.~2006, \aj, 131, 1163

\bibitem[Southworth et al.(2007)]{Southworth:07}
 Southworth, J., Wheatley, P. J., \& Sams,G. 2007, \mnras, 379, 11

\bibitem[Sozzetti et al.(2007)]{Sozzetti:07}
 Sozzetti, A., Torres, G., Charbonneau, D., Latham, D. W., Holman,
 M. J., Winn, J. N., Laird, J. B., \& O'Donovan, F. T. 2007,
 \apj, 664, 1190

\bibitem[Torres et al.(2005)]{Torres:05}
 Torres, G., Konacki, M., Sasselov, D.~D., \& Jha, S. 2005, \apj, 619, 558

\bibitem[Torres et al.(2007)]{Torres:07}
 Torres, G., Bakos, G.~\'A., Kov\'acs, G., Latham, D. W., Fern\'andez,
 J. M., Noyes, R. W., Esquerdo, G. A., Sozzetti, A., et al. 2007,
 \apjl, 666, L121

\bibitem[Torres, Winn \& Holman(2008)]{Torres:08}
 Torres, G., Winn, J. N., \& Holman, M. J.~2008, \apj, 677, 1324

\bibitem[Valenti \& Fischer(2005)]{Valenti:05}
 Valenti, J.~A., \& Fischer, D.~A. 2005, \apjs, 159, 141

\bibitem[Valenti \& Piskunov(1996)]{Valenti:96}
 Valenti, J.~A., \& Piskunov, N. 1996, \aaps, 118, 595

\bibitem[Vogt et al.(1994)]{Vogt:94}
 Vogt, S.~S., Allen, S. L., Bigelow, B. C., Bresee, L., Brown, B.,
 Cantrall, T., Conrad, A., Couture, M., et al.~1994, Proc.~SPIE,
 2198, 362

\bibitem[West et al.(2008)]{West:08}
 West, R. G., Collier Cameron, A., Hebb, L., Joshi, Y. C., Pollacco,
 D., Simpson, E., Skillen, I., Stempels, H. C., et al. 2008, \aap
 submitted, arXiv:0809.4597

\bibitem[Winn et al.(2007)]{Winn:07}
 Winn, J. N., Holman, M. J., Bakos, G. \'A, P\'al, A., Johnson, J. A.,
 Williams, P. K. G., Shporer, A., Mazeh, T., et al. 2007, \aj, 134, 
 1707

\bibitem[Yi et al.(2001)]{Yi:01}
 Yi, S.~K., Demarque, P., Kim, Y.-C., Lee, Y.-W., Ree, C. H., Lejeune,
 T., \& Barnes, S. 2001, \apjs, 136, 417

\bibitem[Zahn(1989)]{Zahn:89}
 Zahn, J.-P.
 1989, \aap, 220, 112

\bibitem[Zucker \& Mazeh(1994)]{Zucker:94}
 Zucker, S, \& Mazeh, T. 1994, \apj, 420, 806

\end{thebibliography}
\end{document}